\documentclass[fleqn,usenatbib]{mnras}

\usepackage{newtxtext,newtxmath}

\usepackage[T1]{fontenc}

\DeclareRobustCommand{\VAN}[3]{#2}
\let\VANthebibliography\thebibliography
\def\thebibliography{\DeclareRobustCommand{\VAN}[3]{##3}\VANthebibliography}

\usepackage{graphicx}	
\usepackage{amsmath}	
\usepackage{subcaption}
\usepackage{orcidlink}
\usepackage{booktabs}
\usepackage{gensymb}
\usepackage{xcolor}
\usepackage{soul}

\newcommand{\kepler}{{\it Kepler}}

\newcommand{\TESS}{{\it TESS}}
\newcommand{\tess}{{\it TESS}}

\newcommand{\plato}{{\it PLATO}}
\newcommand{\gaia}{{\it Gaia}}

\newcommand{\ngts}{{\it NGTS}}

\newcommand{\Ktwo}{K2}

\newcommand{\cheops}{{\it CHEOPS}}

\newcommand{\rpl}{\mbox{R\textsubscript{p}}}
\newcommand{\mstar}{\mbox{M$_{\star}$}}
\newcommand{\rstar}{\mbox{R$_{\star}$}}

\newcommand{\rsun}{\mbox{R$_{\odot}$}}

\newcommand{\rearth}{R$_{\oplus}$}

\newcommand{\teff}{T\textsubscript{eff}}

\newcommand{\tiara}{\texttt{TIaRA}}

\title[TIaRA \tess\ 1]{TIaRA \tess\ 1: Estimating exoplanet yields from Year 1 and Year 3 SPOC lightcurves}

\author[T. Rodel et al.]{
Toby Rodel$^{,1,2,3}$ \thanks{E-mail: trodel01@qub.ac.uk}\orcidlink{0009-0009-2175-7284},
Daniel Bayliss$^{1,2}$ \orcidlink{0000-0001-6023-1335}, 
Samuel Gill$^{1,2}$ \orcidlink{0000-0002-4259-0155}, 
Faith Hawthorn$^{1,2}$ \orcidlink{0000-0002-8675-182X}
\\
$^{1}$ Department of Physics, University of Warwick, Gibbet Hill Road, Coventry, CV4 7AL, UK \\
$^{2}$ Centre for Exoplanets and Habitability, University of Warwick, Gibbet Hill Road, Coventry CV4 7AL, UK \\
$^{3}$ Astrophysics Research Centre, School of Mathematics and Physics, Queen’s University Belfast, Belfast, BT7 1NN, UK \\
}

\pubyear{2023}

\begin{document}
\label{firstpage}
\pagerange{\pageref{firstpage}--\pageref{lastpage}}
\maketitle

\begin{abstract}
We present a study of the detection efficiency for the \tess\ mission, focusing on the yield of longer-period transiting exoplanets ($P > 25$\, days). We created the Transit Investigation and Recoverability Application (\tiara) pipeline to use real \tess\ data with injected transits to create sensitivity maps which we combine with occurrence rates derived from \kepler.  This allows us to predict longer-period exoplanet yields, which will help design follow-up photometric and spectroscopic programs, such as the \ngts\ Monotransit Program. For the \tess\ Year 1 and Year 3 SPOC FFI lightucurves, we find $2271^{+241}_{-138}$ exoplanets should be detectable around AFGKM dwarf host stars. We find $215^{+37}_{-23}$ exoplanets should be detected from single-transit events or "monotransits". An additional $113^{+22}_{-13}$ detections should result from "biennial duotransit" events with one transit in Year 1 and a second in Year 3. We also find that K dwarf stars yield the most detections by \tess\ per star observed. When comparing our results to the TOI catalogue we find our predictions agree within $1\sigma$ of the number of discovered systems with periods between 0.78 and 6.25\,days and agree to $2\sigma$ for periods between 6.25 and 25\,days. Beyond periods of 25\,days we predict $403^{+64}_{-38}$ detections, which is 3 times as many detections as there are in the TOI catalogue with $>3\sigma$ confidence. This indicates a significant number of long-period planets yet to be discovered from \tess\ data as monotransits or biennial duotransits. 
\end{abstract}




\section{Introduction}
\label{section:Intro}
Transiting exoplanets are of exceptional scientific importance as they allow for many parameters of a planetary system to be characterised. Transit detections themselves allow for the orbital period and radius ratio of a planet and its host to be constrained \citep{2014Winn}. If the host star is sufficiently bright, spectroscopic radial velocity measurements can be used to constrain the planetary mass. These can be combined with transit radius measurements to constrain the planetary density and make inferences about composition and internal structure.  The brighest host stars also permit atmospheric characterisation of planets via transmission spectroscopy \citep[e.g.][]{2002Charbonneau, 2014Madhusudhan, 2018Kreidberg, 2018Kempton, 2019Madhusudhan}.

Discoveries from transit surveys are biased towards shorter-period planets.  Shorter-period planets have a greater geometric probability of transit \citep{2014Winn} and transit more frequently within any given monitoring campaign, resulting in an increase in the signal-to-noise ratio (SNR) of the transit signal. Despite this longer-period planet discoveries are valuable.

Longer-period planets yield insights into planet formation and migration mechanisms. At the extremely small orbital separation of planets with $P<10$\,days star-planet interactions may have a significant effect on dynamical evolution \citep{2015Valsecchi}. This leads to very reduced eccentricities \citep{2012Albrecht, 2014Winn}, effectively removing any trace of past dynamical interactions from the planet's present day orbit. Planets with longer-orbital periods have larger orbital separations and as such experience weaker star-planet interactions, retaining more information on past dynamical interactions. Giant planets on extremely short periods (i.e. "hot-Jupiters") and those on longer periods ("warm-Jupiters") have been shown to exhibit differences in their distributions of orbital eccentricity \citep{2021Dong} and the number of close orbital companions \citep{2016Huang}. This indicates that giant planets with orbital periods longer than 10 days may experience different migration pathways compared with short-period giant planets \citep{2011Wu, 2014Petrovich, 20170Mustill}. Furthermore, Rossiter-McLaughlin \citep[RM;][]{rossiter1924, mclaughlin1924} measurements of orbital obliquity allow further constraining of possible migration mechanisms in the planet's history \citep{Triaud2018}. Already, \cite{2022Rice} have found a trend towards more aligned obliquities in some warm-Jupiters compared to hot-Jupiters lthough the number of well-studied warm-Jupiters is comparatively fewer.

Additionally, short-period planets experience a greater level of atmospheric mass loss than those on longer-periods \citep{2019Owen}, meaning longer period planets will retain more of their atmospheres. This makes long-period planets promising targets for atmospheric studies to probe cooler atmospheres that have undergone less photo-evaporation and to gain an accurate understanding of their formation \citep{2011Oberg, 2014Madhu}. However, in spite of their promise, longer-period cooler planets are more difficult targets for transmission spectroscopy than closer-in hotter planets due to their generally smaller atmospheric scale height \citep{2018Kempton}.

Long-period planets also offer important targets in the search for exomoons and exorings, which have not yet been detected but are theoretically detectable around giant transiting planets \citep{2004Barnes, 2009Kipping, 2012Simon, 2017Aizawa}. Exomoons and exorings are predicted to be more stable and thus likelier to exist around longer-period planets with greater orbital separations \citep{2002Barnes, 2004Barnes, 2009Cassidy, 2021Dobos, 2023Makarov}.

\begin{figure}
    \includegraphics[width=\columnwidth]{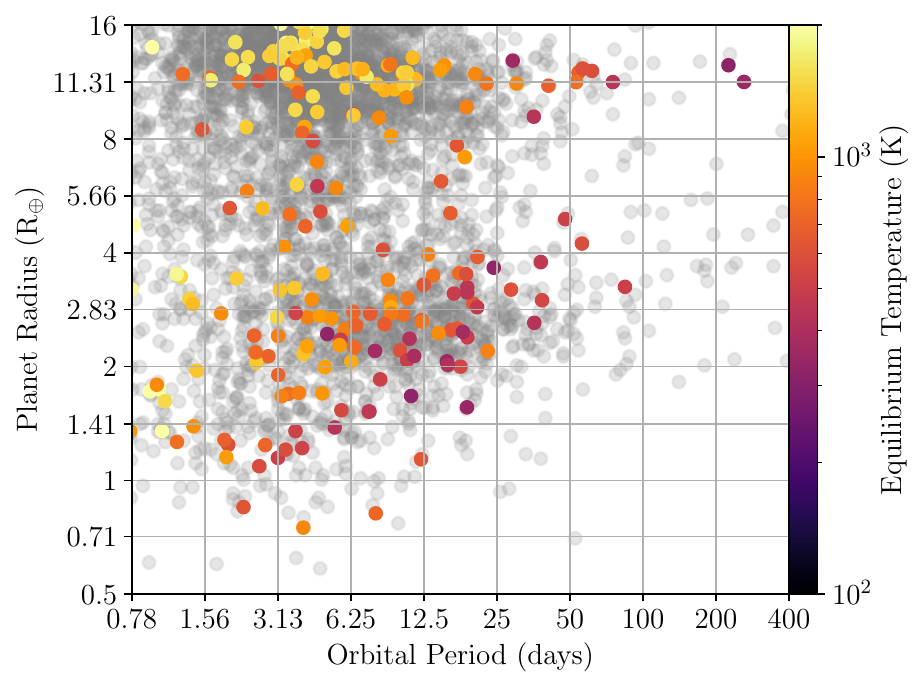}
    \caption[The \tess\ planet sample]{The \tess\ planet sample. Full sample of \tess\ objects of interest (TOI) are shown in grey and confirmed TOIs are coloured according to their equilibrium temperature. The full 5910 TOIs consist of the \tess\ project candidate list excluding those flagged as False Positives.  The 273 confirmed \tess\ planets include all published and confirmed exoplanets from \tess\ with fully constrained, masses, radii and orbital periods. Both data sets were downloaded from the NASA exoplanet archive \citep{2013Akeson} on 2023-11-23.}
    \label{fig: radius-period}
\end{figure}

Since 2018, the Transiting Exoplanet Survey Satellite \citep[\tess; ][]{2014Ricker} has been performing an all-sky survey searching for transiting exoplanets around bright host stars. One of the primary science goals is to find  nearby planets amenable to atmospheric characterisation \citep{2014Ricker, 2018Kempton}. \tess~observes each $24\degree \times\ 96\degree$\ sector of sky for 27\,days at a time, although overlap of some regions between sectors mean that the observation baselines for some targets will be >300\,days near the Ecliptic poles in the Continuous Viewing Zone \citep[CVZ;][]{2014Ricker}.  However, in the first year of \tess\ $\approx75\%$ of target stars were only observed in a single sector which means that longer period planets are likely to only be observed as a single transit. Figure~\ref{fig: radius-period} shows the population of \tess\ planet detections both confirmed and unconfirmed, showing the relative lack of longer-period planets. Although such single-transiting candidates are harder to detect, it is not impossible and the ability of \tess\ to do so has been studied previously in \citet{2018Cooke, 2019Cooke, 2019Villanueva}. Additionally, a number of long-period transiting planets have been discovered via monotransits from the \Ktwo\ \citep{2014Howell} mission \citep[e.g;][]{2015Vanderburg, 2016Vanderburg, 2018Giles, 2019Kruse}.

Planets discovered from a single transit ("monotransits") do not have constrained periods and require follow-up observations to rule out false positive scenarios and constrain their periods. Such observations can be performed using photometry or spectroscopic radial velocity (RV) measurements.

Photometric observations can be used to detect additional transit events, which will either determine the period uniquely, or provide a discrete set of aliases \citep[e.g;][]{2020Gill}.  This is one of the science goals of the Next Generation Transit Survey \citep[\ngts;][]{2018Wheatley} monotransit working group, which uses the 12-telescope \ngts\ facility in Paranal, Chile to monitor \tess\ monotransit candidates that display transit events with depths 1000\,ppm or greater \citep{2020Bayliss}. This program has already successfully confirmed  "monotransits" from \tess~data \citep[e.g.][]{2020Gill, 2020Lendl, 2022Ulmer-Moll, 2022Grieves}.  The TESS single transit planet group has also been following up monotransits using a global network of small telescopes \citep{2023Dragomir}.

For shallower \tess\ monotransits ($<$1000\,ppm) usually associated with smaller planets it is extremely difficult to detect additional transits from the ground.  In these cases space facilities such as \cheops\ \citep{2021Benz} can be used to detect further transits, e.g. \citep{2023Tuson, 2023Osborn, 2023Ulmer-Moll}.  

Spectroscopic monitoring of monotransits can also determine the true period of a planet candidate and rule out some eclipsing binary stars which display very large RV variations \citep[e.g;][]{2019Dragomir, 2022Ulmer-Moll, 2022Eberhardt}. However, this requires long-term spectroscopic monitoring of candidates with high resolution instruments. Such spectrographs have to be mounted on large telescopes for which observing time is limited.

Given the large-effort and hundreds of hours of telescope time involved with such follow-up efforts, it is vital to inform these efforts with up-to-date understanding of the sensitivity of \tess\ to monotransit events and the expected yield of monotransits from the mission.

Previous studies of yields from the \tess~mission have largely focused on multi-event signals. Prior to the launch of \tess, yield estimates were predicted in \citet{2015Sullivan, 2017Bouma} based on the expected performance of the mission given in \cite{2014Ricker}. 

Around the time that \tess\ was launched there were several studies that estimated the potential yields from the \tess\ mission, including: \citet{2018Muirhead, 2018Huang, 2018Barclay, 2019Ballard}. These studies were still largely based on the expected performance of the \tess\ mission as the data from the mission was only just becoming available.  More recently, \citet{2022Kunimoto} have revised the yield estimates based on the performance of \tess\ in its primary and first extended mission.  All of these studies were largely focused on planet discoveries from two or more transits.

The studies of \citet{2018Cooke} and \citet{2019Villanueva} were specifically focused on the expected yield of monotransits from \tess. These studies were mostly based on the expected performance of \tess\ from \cite{2014Ricker} due to the relative lack of \tess\ data available at the time of publication shortly after \tess\ had launched. \cite{2019Cooke} provided an update to the yields from \cite{2018Cooke} using the performance of the \tess\ primary mission and estimating the yield when \tess\ revisited the southern ecliptic in year 3 during its first extended mission. All of these studies found that a significant number of long-period planets will be detectable in \tess\ lightcurves as monotransits.

In this study, we use the custom-made Transit Investigation and Recoverability Application (\tiara) pipeline to create sensitivity maps using real \tess\ Science Processing Operations Centre \citep[SPOC;][]{2019Stassun} lightcurves from the southern ecliptic hemisphere (years 1 and 3). We combine these with occurrence rates to estimate a yield of planet discoveries.

We present the full details of the \tiara\ pipeline in Section~\ref{section:tiara} of this paper.  In Section~\ref{section:tess_method} we describe the application of the \tiara\ pipeline to the SPOC FFI lightcurves from Year 1 and Year 3 of the \tess\ mission.  In Section~\ref{section:results} we present sensitivity maps and estimates of exoplanet yields from our simulation. In Section~\ref{section:discussion} we discuss interpretations of our results and some potential caveats and compare them to both actual \tess\ discoveries and other yield predictions. Finally in Section~\ref{section:conclusion} we summarise our work and discuss the potential for applying \tiara\ to additional \tess\ data sets as well as other transit surveys such as \plato\ \citep{2014Rauer}.

\section{The \tiara\ pipeline}
\label{section:tiara}
In order to estimate the discovery yield of planets from transit surveys such as \tess, we need to simulate as many transiting exoplanet signals as possible in a realistic manner.  To do this, we base our simulation on the actual stars that are monitored in the survey.  We use the timestamps, measured noise properties, and dilution factors that are recorded for each star.  We also use all available stellar properties such as radius (\rstar), temperature (\teff), and mass (\mstar) - which are informed by \gaia\ data releases \citep{2016Gaia, 2018Gaia, 2023Gaia}.

In this Section we introduce the Transit Investigation and Recoverability Application (\tiara) pipeline, which has been developed to calculate the sensitivity of a given transit survey to discovering transiting planets across a range of orbital periods and planetary radii. When combined with occurrence rates, this provides yield statistics for surveys that can be used to understand and assess the completeness of discoveries from a survey and plan future surveys.

\subsection{Input of Lightcurves}
\label{subsection:ingesting}
\tiara\ requires a catalogue of stars observed by a survey and their parameters. Using these target lists we retrieve the relevant lightcurves for processing. We use the timestamps in the lightcurve to calculate a window function for each target star and to determine the number of in-transit datapoints when calculating the signal-to-noise (SNR) for transit events. We also use measurements of noise values and dilution ratios for each lightcurve to calculate SNR for transit events. Using noise values calculated from real lightcurves allows for us to realistically account for a range of astrophysical and instrumental effects in the data.  Similarly we use the dilution ratio to account for blending of sources which is a very common problem in wide field photometric transit surveys.

\subsection{Simulating Transiting Planets}
\label{subsection:generate}

\tiara\ simulates transiting exoplanets by generating planet parameters and then calculates which timestamps are either occurring in the ingress, egress or fully in-transit portion of the transit event.  

\subsubsection{Generation of planetary parameters}
\label{subsubsection:planet_params}

Thanks largely to the \kepler\ mission \citep{2010Borucki}, we have a good understanding of the occurrence rates of exoplanets, particularly for those with orbital periods less than $<$100 days \citep{2013Fressin, 2013Dressing, 2015Dressing, 2019Hsu, 2020Kunimoto}.  For example we know that super-Earth-sized exoplanets are far more common than gas giant planets in short period orbits.  We can take advantage of this understanding to optimize \tiara\ by simulating transiting planets in proportion to the occurrence rate of those planets. The occurrence rates we used are described in Section~\ref{subsection:occurance_rates}. Note that in including planets in the simulation proportionally to occurrence rates is \textit{not} an attempt to simulate the underlying planetary population at this stage but instead is a measure to optimize the simulation. We prioritise injection of more numerous types of planets to avoid wasting simulation time on calculating a precise detection efficiency for rare types of planets for which the yield calculation will almost certainly be zero due to the small numbers in the underlying population. This means that the sensitivity maps we produce are still population agnostic, just with less precise values for rarer types of exoplanets.

We simulate a large number of planets per star ($N_\text{Pl}$) to obtain a more robust simulation. To choose the orbital period and radius of each of these we first select a period-radius bin using the occurrence rates as a weighted probability for a random draw. The exact period and radius of each planet in the sample was taken from a uniform distribution between the upper and lower limits of the chosen period and radius bins. We use the generated period and the mass of the star to estimate the semi-major axis using Kepler's third law under the assumption that the mass of the planet is negligible compared to that of the star.

In order to ensure our simulated planets have a realistic eccentricity distribution, we randomly assign each planet an orbital eccentricity from a beta distribution following the prescription set out in \citet{2013Kipping}.  We adopt the values of $\alpha=1.03$ and $\beta=13.6$ as proposed by \citet{2015VanEylen}.

We randomly assign a periastron angle ($\omega$) in radians, from a uniform distribution over the full $\pi$ radian range.

For each planet generated, we calculated the geometric probability of transit ($p_{\text{tran}}$) following \citet{2014Winn}:

\begin{equation}
    p_\text{tran} = \left(\frac{\rstar+\rpl}{a}\right)\left(\frac{1+e\sin\omega}{1-e^2}\right).
    \label{eq: ProbTransit}
\end{equation}

For each simulated transiting planet we randomly generate $N_b$ different impact parameters $(b)$ from a uniform distribution between the values of 0 and $1+\frac{\rpl}{\rstar}$.

\subsubsection{Simulation of transits}
\label{subsubsection:transit_timestamp}

We calculate the total transit duration $(T_{14})$ including the ingress and egress between the first and fourth points of intersection between the planetary and stellar discs using the following equation from \citet{2014Winn}:

\begin{equation}
    T_{14} = \frac{P}{\pi}\arcsin\left[\frac{\rstar}{a}\frac{\sqrt{\left(1+\frac{\rpl}{\rstar}\right)^2-b^2}}{\sin i}\right]\frac{\sqrt{1-e^2}}{1+e\sin\omega},
    \label{eq: duration_14}
\end{equation}

where $i$ is the inclination of the orbital plane and $a$ is the semi-major axis of the orbit.

In addition, we also calculate the duration of transit between the second and third intersection $(T_{23})$ using the following equation from \citet{2014Winn}:

\begin{equation}
    T_{23} = \frac{P}{\pi}\arcsin\left[\frac{\rstar}{a}\frac{\sqrt{\left(1-\frac{\rpl}{\rstar}\right)^2-b^2}}{\sin i}\right]\frac{\sqrt{1-e^2}}{1+e\sin\omega}.
    \label{eq: duration_23}
\end{equation}

For grazing transits this does not give real solutions, and in such cases we simply set the value of $T_{23} = 0$.

To save computation time, we do not simulate signals with epochs that would result with zero in-transit data points. To accomplish this we convert the timestamps ingested by \tiara\ into orbital phase using a chosen reference time  (TBJD=0, BJD=2457000 for \tess) and use these phase arrays to determine which portions of the planets orbit are monitored by \tess. We then use these observed stretches of phase to generate random epochs in the form of an offset to the chosen reference time, expressed in orbital phase. These observed phase arrays also allow us to calculate the probability that any transit would randomly fall within an observed timespan of \tess\ ($p_\text{obs}$). We generate $N_\text{ph}$ values of this phase offset for each inclination of each planet we simulate resulting in $N_b\times N_\text{ph}$ total transit signals per planet. 

Using the values of $T_{14}$ and $T_{23}$ calculated from Equations~\ref{eq: duration_14}~$\&$~\ref{eq: duration_23} and the epoch of each signal, we count the number of points in the timestamps of each lightcurve which are in ingress ($N_\text{ingress}$), egress ($N_\text{egress}$) or full transit ($N_\text{full}$). To obtain a more realistic value of the signal to noise we use a trapezoidal approximation for the shape of transit, and if the signal to noise calculation is likened to the effective area of the transit curve, then the ingress and the egress are triangular while the full transit is rectangular. Thus the effective count of in-transit points for a trapezoidal transit is:

\begin{equation}
    N_\text{trans, eff}=N_\text{full}+0.5(N_\text{ingress}+N_\text{egress}).
    \label{eq:in-trans_points}
\end{equation}

It is worth noting that this trapezoidal model is a simplification of the actual shape of a transit and does not account for effects such as limb darkening\citep{2016Espinoza, 2020Agol}. We believe this simplification is still robust enough to produce good yield estimates and discuss these effects further in Section~\ref{subsection:trapezoid-discussion}

\subsection{Signal-to-Noise Calculation}
\label{subsection:recovery}
A full injection and recovery test \citep[e.g;][]{2015Dressing, 2019Hippke, 2023Bryant} would require additional computation time to run a detection algorithm \citep[e.g. BLS;][]{2002Kovacs, 2006Collier} on the data and to initialise a full transit model such as those produced by the \texttt{batman} package \citep{2015Kriedberg}. We instead calculate detection based purely on the SNR for each signal based on the generated planet properties and lightcurve properties.

The width of the transit is equal to the effective number of in-transit points from Equation~\ref{eq:in-trans_points} multiplied by the time cadence of the lightcurve $(\Delta T)$ For non-grazing transits where $0\leq b<1-\frac{\rpl}{\rstar})$ the depth of transit is simply:

\begin{equation}
    \delta=\left(\frac{\rpl}{\rstar}\right)^2.
    \label{eq:depth_full}
\end{equation}

For grazing transits $(1-\frac{\rpl}{\rstar}\leq b \leq 1+\frac{\rpl}{\rstar})$ we calculate the true anomaly at the time of mid transit in accordance with \citet{2016Maxted}. We then use equation 5.63 from \citet{hilditch_2001} to calculate the corresponding projected orbital separation. We then calculate the depth of transit as the overlap in area between the planetary and stellar disks at this projected separation.

In either case, we then calculate SNR using the equation below:

\begin{equation}
    \text{SNR} = \frac{\delta}{1+C}\frac{\sqrt{N_\text{trans, eff}\Delta T}}{\sigma},
    \label{eq: SN}
\end{equation}

where the contamination ratio $(C)$ is the proportion of flux from background objects rather than the target itself, $\Delta T$ is the time cadence of the lightcurve and $\sigma$ is the lightcurve noise on the timescale of $\Delta T$.  This allows us to account for changes in cadence of observations such as the change from 30 to 10 minute cadence between Year 1 and Year 3 of \tess\ full frame images.

\subsection{Detection Probability}
\label{subsection:threshold}
To convert from SNR to a recovery rate, we need a function that encapsulates the likelihood that a given signal with a certain SNR will be detected as a transit candidate. Previous studies \citep{2015Sullivan, 2017Bouma, 2018Huang, 2018Barclay, 2018Cooke, 2019Villanueva} have used an SNR threshold of $\text{SNR}\geq7.3$ to determine whether a planet is detectable. This approach is essentially a step function, where the probability of detecting a transit at $\text{SNR}\geq7.3$ is unity and the probability of detecting a transit at $\text{SNR}<7.3$ is zero. \cite{2022Kunimoto} used an incomplete gamma function ($\gamma$) initially fitted to the \kepler\ pipeline by \cite{2017Christiansen, 2019Hsu}, to characterise the probability of a planet being both detected and passing vetting. The form of this function is:

\begin{equation}
    p_\gamma\left(\text{det}\right) = c_{N_\text{tr}}\times\gamma\left(\alpha_{N_\text{tr}}, \frac{\text{S/N}}{\beta_{N_\text{tr}}}\right)
    \label{eq: gamma}
\end{equation}

where $c_{N_\text{tr}}$ is the maximum probability of detection a transit signal with $N_\text{tr}$ observed transits can reach and $\alpha_{N_\text{tr}},~\beta_{N_\text{tr}}$ are coefficients determined from the value of $N_\text{tr}$ using Table~\ref{table:gamma_coeffs}.

For determining a detection probability, \tiara\ uses the \kepler\ gamma functions set out in \cite{2017Christiansen} and as adopted by \cite{2019Hsu}.  For testing and comparison purposes, \tiara\ can also used a fixed detection threshold such as $\text{SNR}\geq7.3$.

\begin{figure}
    \centering
    \includegraphics[width=\columnwidth]{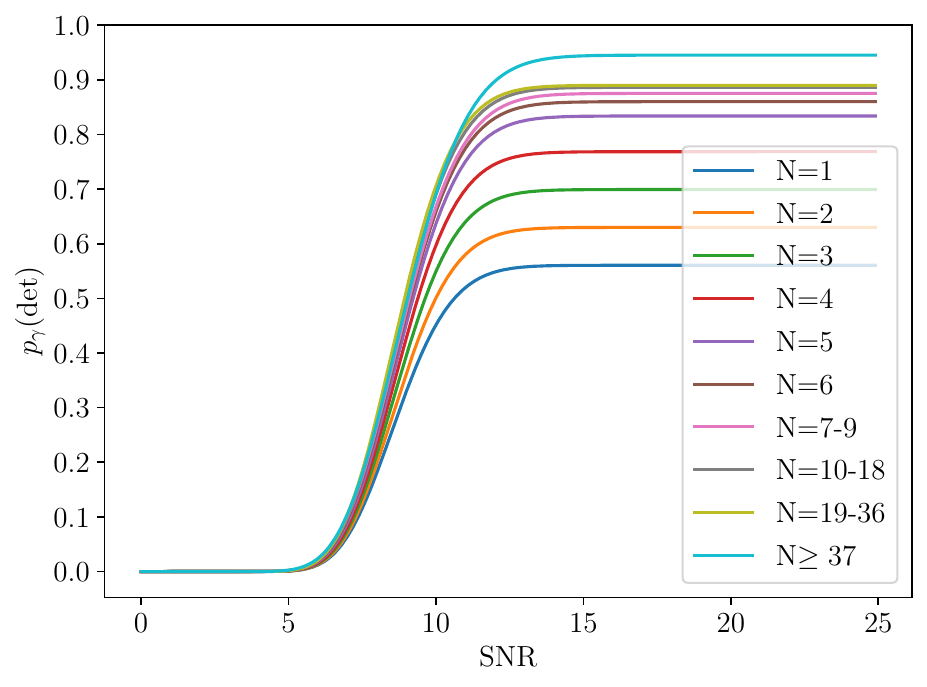}
    \caption[Gamma function plot]{The set of incomplete gamma functions used as detection probability functions in the \tiara\ pipeline.  Each function shows the probability a planet will be detected given the signal-to-noise (SNR) of its transit signal in the lightcurve.  Each function is for a different number of individual transit events in the lightcurve.  The monotransits and duotransits are represented by the blue N=1 and orange N=2 functions respectively.}
    \label{fig:gamma}
\end{figure}

In \cite{2019Hsu}there is no detection probability function defined for the case of only one or two transit events in the lightcurve. However calculating the yields for monotransits and duotransits is a key aspect of the \tiara\ pipeline.  Therefore we perform a linear extrapolation of the coefficients fitted by \cite{2019Hsu} to obtain values for $N_\text{tr}=1$ and $N_\text{tr}=2$ resulting in Table~\ref{table:gamma_coeffs}. The incomplete gamma function is represented graphically for each set of coefficients in Figure~\ref{fig:gamma}.

Applying a gamma function based on the detection efficiency of \kepler\ to other surveys such as \tess\ comes with caveats as different detection pipelines will have different performance for the same signal to noise value. The effects of this on our yield predictions for \tess\ is discussed further in Section~\ref{subsection:p_det}

\begin{table}
\caption[Gamma function coefficients]{Gamma function coefficients.}
\centering
\begin{tabular}{lllll}
\hline
$N_\text{tr}$ & $\alpha_{N_\text{tr}}$ & $\beta_{N_\text{tr}}$ & $c_{N_\text{tr}}$ & Source                          \\
\hline
1             & 34.3932                & 0.254262              & 0.560547          & This work                       \\
2             & 33.8908                & 0.259367              & 0.629820          & This work                       \\
3             & 33.3884                & 0.264472              & 0.699093          & \cite{2019Hsu}                  \\
4             & 32.8860                & 0.269577              & 0.768366          & \cite{2019Hsu}                  \\
5             & 31.5196                & 0.282741              & 0.833673          & \cite{2019Hsu}                  \\
6             & 30.9919                & 0.286979              & 0.859865          & \cite{2019Hsu}                  \\
7-9           & 30.1906                & 0.294688              & 0.875042          & \cite{2019Hsu}                  \\
10-18         & 31.6342                & 0.279425              & 0.886144          & \cite{2019Hsu}                  \\
19-36         & 32.6448                & 0.268898              & 0.889724          & \cite{2019Hsu}                  \\
$\geq37$      & 27.8185                & 0.32432               & 0.945075          & \cite{2019Hsu}
\end{tabular}
\label{table:gamma_coeffs}
\end{table}

\subsection{Minimum detectable radius cutoff}
\label{subsection: radius_cut}
Since we inject planets proportionally to their real occurrence rates, we will simulate many more smaller radius planets than larger planets due to their relative occurrence rates.  However many of the smaller planets would produce transits with extremely low SNR, which may be far below the level of detectability in a given survey.  Therefore to conserve computational resources and increase the efficiency of the simulation we implement a radius cutoff to the occurrence rates based on the precision of the given survey. To do this we estimate the minimum detectable radius from a single transit.

To calculate the minimum detectable radius we assume a favourable set of transit parameters which consist of $e=0$, $b=0$, and $T_\text{dur}$ set to a "long" transit duration - defined to be the duration of a P=1000\,days transiting exoplanet of negligible radius around the given host star. Since SNR increases with transit duration, this long duration - beyond what we expect to realistically transit, was chosen so that small-radius planets which could be detected from a long enough transit were not removed from the simulation. 

We choose a SNR threshold of 4 to determine the minimum detectable radius as this correlates with a very low probability of detection: $<10^{-5}$ (see Section~\ref{subsection:threshold}).For planets with radii below the cutoff, we simply assign a probability of detection equal to 0.

\section{\tess\ simulations}
\label{section:tess_method}
\subsection{\tess\ SPOC FFI lightcurves}
\label{subsection: stellar method}
In this study we apply \tiara\  to the Year 1 and Year 3 Full Frame Image \citep[FFI;][]{2014Ricker} lightcurves from the \tess\ Science Processing Operations Center \citep[SPOC;][]{2016Jenkins}. The SPOC FFI lightcurve sample is a high quality, homogeneous and readily availiable dataset representing \tess\ targets which are most amenable to exoplanet detection and thus is appropriate for our purposes \citep{2020Caldwell}. Additionally the \ngts\ monotransit working group uses the SPOC FFI lightcurves in their search for \tess\ monotransits \citep[e.g;][]{2020Gill} which allows for easy comparison of our results. We compiled the SPOC FFI target lists per \tess\ sector from MAST availiable at \url{https://archive.stsci.edu/hlsp/tess-spoc} into a single list containing the \tess~Input Catalogue \citep[TICv8;][]{2019Stassun} numbers of all the SPOC FFI target stars and the \tess\ sectors they were observed in. We also removed all stars with radii $<0.1$\rsun\ and $>4$\rsun\ from the sample. This results in a sample of 1323228 stars across Year 1 and Year 3. For every star in this list, we obtained the relevant lightcurve \texttt{FITS} files via the MAST portal at \url{https://archive.stsci.edu/}.  

\subsection{Stellar parameters}
\label{subsection:parameters}
We extracted the stellar radius (\rstar), \tess\ magnitude $(T_\text{mag})$, and effective temperature (\teff) directly from the \texttt{FITS} file headers: These stellar parameters are themselves sourced from the TICv8 \citep[see][]{2019Stassun}. 

To calculate a semi-major axis for our injected planets, we require a stellar mass (\mstar). Due to a large number of targets lacking a value of the stellar mass in the lightcurve \texttt{FITS} file headers, we estimate the stellar mass using the following power law, which is approximated from the mass and radius of binary stars \citep{1988Harmanec}:

\begin{equation}
  \mstar=\rstar^{1.25},  
\end{equation}

where \mstar\ and \rstar\ are both in solar units. The dependence of the semi-major axis on stellar mass is relatively weak and thus has a minor effect on planet detectability, therefore this power law is adequate for the main-sequence dwarf stars in the SPOC FFI sample.  We assign a spectral type to each star based on \teff\ using definitions from \cite{2013Pecaut}. Stars with \teff\ hotter than the maximum cutoff for A type stars (10050\,K) were marked as spectral type OB.

\subsection{Window functions}
\label{subsection: Coverage}
In order to use the real window function for each star in the SPOC FFI list, we determine blocks of continuous \tess\ observations for each target using the timestamps from the \texttt{FITS} file \texttt{TIME} header for each \tess\ sector.  We only consider good quality data where the data quality flag (\texttt{QUALITY}) = 0.  We use the first and last timestamps of each Sector to define the block of continuous data for each Sector. To identify gaps within Sectors, we search for instances where the difference between two consecutive  timestamps is greater than 0.5\,days. Most of these gaps are the mid-Sector gaps, which occur during the perigee of the 13.7\,day \tess\ orbit when data is downloaded to ground-stations \citep{2014Ricker}.  Other data gaps are caused by technical issues with the specific \tess\ camera, the entire \tess\ spacecraft, or the variety of reasons that give rise to non-zero data quality flags (e.g. stray light,cosmic rays, spacecraft momentum dumps and pointing issues).

\begin{figure*}
    \centering
    \includegraphics[width=\textwidth]{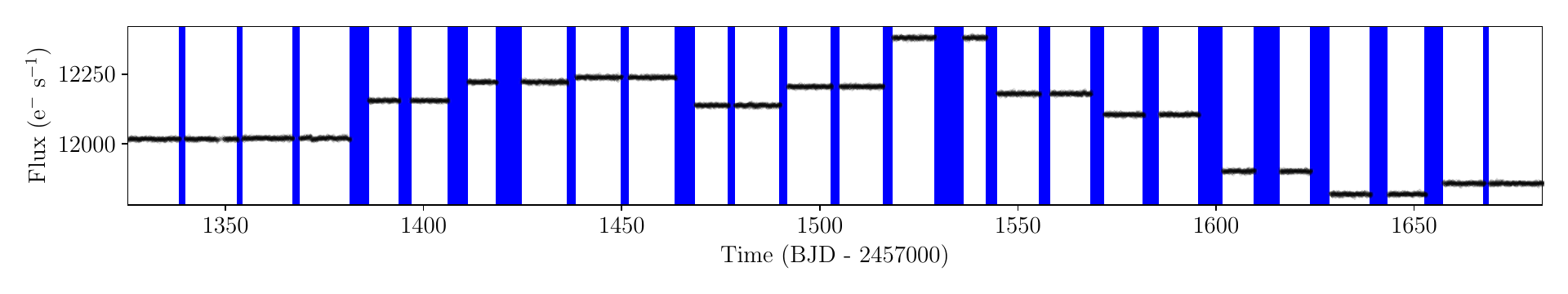}
    \caption[\tess\ CVZ lightcurve]{An example SPOC FFI lightcurve from a star (TIC-261236954) in the CVZ of \tess\ in Year 1 of the mission, illustrating gaps in the data.  The blue shaded regions show data gaps longer than 0.5\,days.  In this example we find that over the course of one year 79.3\% of time is photometrically monitored, while the gaps account for 20.7\% of the year.}
    \label{fig: limlc}
\end{figure*}

To illustrate these data gaps, we plot the lightcurve of a typical SPOC FFI target (TIC-261236954) in Figure~\ref{fig: limlc}.  TIC-261236954 is in the \tess\ CVZ, so can be used to illustrate gaps over the duration of Year 1 of the \tess\ mission. For this star the blocks of continuous photometry account for 79.3\% of the total Year 1 duration, while the gaps make up 20.78\% of that year.

\begin{figure*}
    \centering
    \begin{subfigure}[t]{\columnwidth}
        \centering
        \includegraphics[width=\columnwidth]{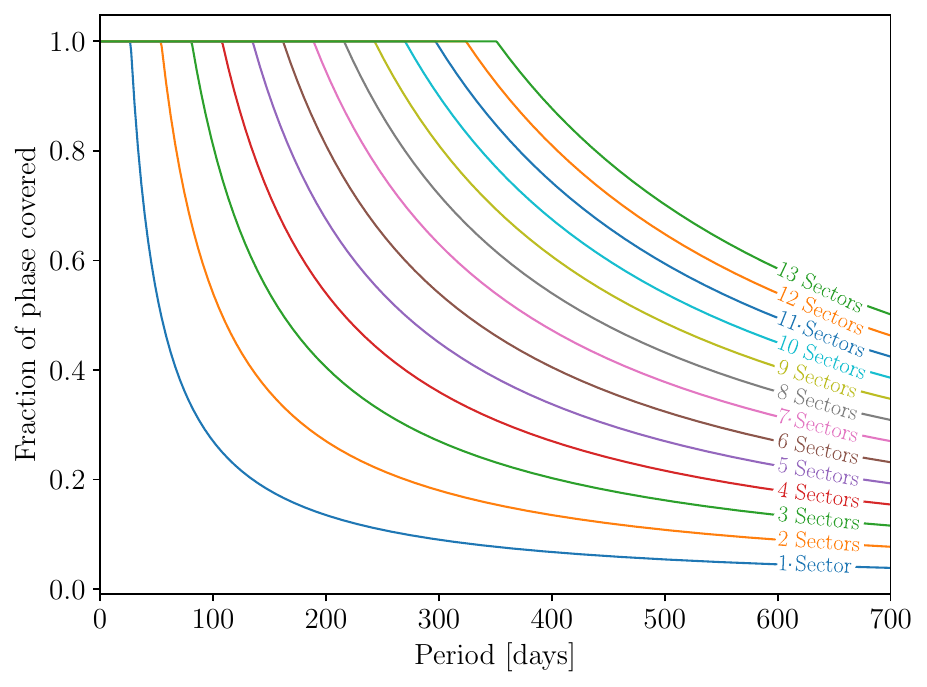}
    \end{subfigure}
    \begin{subfigure}[t]{\columnwidth}
        \centering
        \includegraphics[width=\columnwidth]{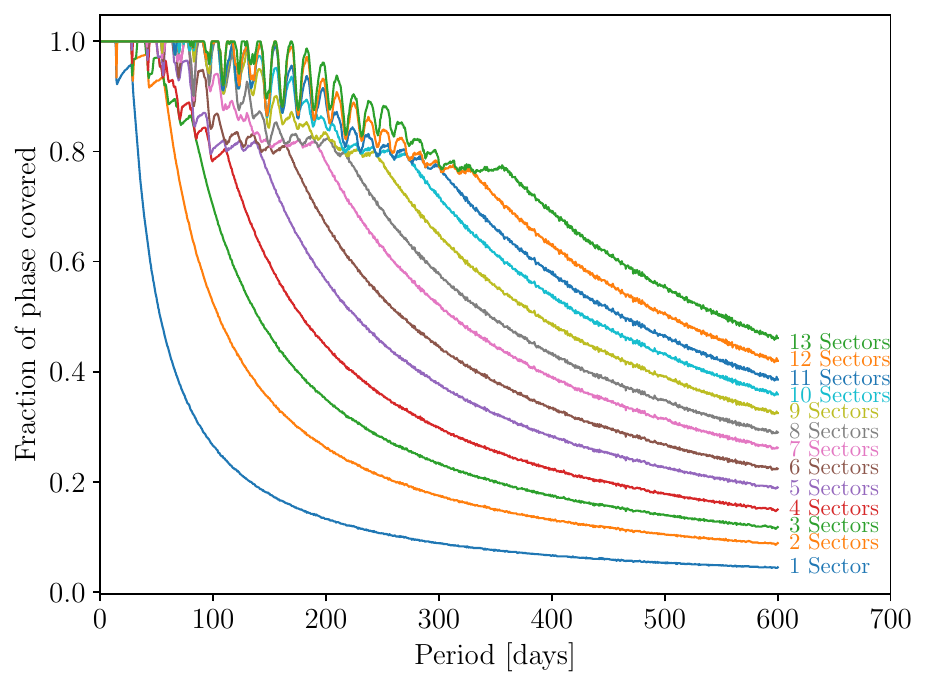}
    \end{subfigure}
    \caption[Comparison of \tess\ window functions]{Completeness of phase coverage as a function of planetary orbital period.  \textit{Left:} Idealised 27\,day coverage (no gaps) for durations ranging from 1 to 13 consecutive sectors.  \textit{Right:} The real phase coverage function for an example SPOC FFI target (TIC-261236954) in the \tess\ CVZ.}
    \label{fig: coverage plot}
\end{figure*}

In order to demonstrate the effect of these gaps in the \tess\ window functions, we calculate the fraction of orbital phase covered by \tess\ observations as a function of the orbital period of a simulated planet for two scenarios:  (1) an idealised 27\,day window function, and (2) the window function of TIC-261236954 (the lightcurve for which is set out in Fig~\ref{fig: limlc}. This allows us to compare the effect of gaps for up to 13 consecutive Sectors.  The results are set out in Figure~\ref{fig: coverage plot}, and show the significant difference between the idealised 27\,day Sectors and the real TIC-261236954 lightcurve window function. This highlights the need to account for the gaps in the \tess\ data for each star as we do in the \tiara\ pipeline.  We note that different pipelines may use different data quality flags, which may slightly alter the position and/or duration of some of the data gaps.

\subsection{Signal to Noise}
\label{subsection: TESS SNR}
To calculate the SNR (Equation~\ref{eq: SN}) for each simulated transiting planet, we take the noise of the lightcurve ($\sigma$) to be the Combined Differential Photometric Precision \citep[CDPP;][]{2012Christiansen} 2hr noise as produced by the SPOC FFI pipelilne and recorded in the \texttt{FITS} headers as \texttt{CDPP2\_0}. This provides us with a readily available pre-calculated noise metric, which saves computation time. Similarly we take the source-to-background brightness ratio $(C+1)$ from the SPOC FFI pipeline recorded in the \texttt{FITS} headers as \texttt{CROWDSAP}. The time cadence $(\Delta T)$ has changed over the course of the \tess\ mission.  For Year 1 the time cadence was 30\,mins, while for Year 3 the time cadence was 10\,mins.

\subsection{Probability of detection}
\label{subsubsection: p_det}
We assume that the detection and vetting of transiting exoplanet signals in \tess\ is similar to that of the \kepler\ mission, and we therefore use the the modified version of the \kepler\ gamma function (Section~\ref{subsection:threshold}) to define our probability of detection for each simulated transiting planet. For comparative purposes, we also record detections via two threshold SNR criteria: at SNR$\geq$7.3 (to match some previous yield studies) and SNR$\geq$20 (to match monotransit searches that typically require such a high SNR for robust detections).

\subsection{Creating sensitivity maps}
\label{subsection: grids binning}
For each transit signal that we generate, we convert the SNR into detection probability as described in Section~\ref{subsubsection: p_det}. Each of these probabilities is then multiplied by the probability of observation for each signal calculated from its window function (see Section~\ref{subsection: Coverage}) to give a probability of observation and detection. These values are again multiplied by the geometric probability of transit for each signal to give a probability of transit, observation and detection by \tess\ for each signal. We take the binned averages for all signals using the same period and radius bins as \cite{2020Kunimoto}.

We treat each of our detection probabilities as a Poisson statistic such that the error in each is equal to its value divided by the square root of the total count of signals in that bin.

\subsection{Occurrence Rates}
\label{subsection:occurance_rates}
In order to convert our sensitivity maps to yield estimates for the Year 1 and Year 3 \tess\ SPOC FFI sample, we need to know the occurrence rates for the underlying population of planets orbiting each type of star in the sample. For F,G and K dwarf stars we use the occurrence rates calculated in  \citet{2020Kunimoto}.  For A type stars we do not have robust occurrence rates, so we use the same grid as for F type stars.

\cite{2020Kunimoto} did not calculate occurrence rates for M dwarfs, so for consistency and ease of comparison between our results we rebin the M dwarf occurrence rate grid from \cite{2015Dressing} to use the same bins as the grids used in \cite{2020Kunimoto}. To do this we use a bivariate spline approximation over a rectangular mesh using the \texttt{RectBivariateSpline} interpolator from the \texttt{python} module; \texttt{scipy.interpolate} \citep{2020SciPy-NMeth}. We take the midpoints of each period and radius bin from the \cite{2015Dressing} grid and use these for the coordinates of each grid value. We then used the median value, lower bound, and upper bound of each grid cell with these coordinates to create three separate interpolators for each value. We then feed the midpoints of all the \cite{2020Kunimoto} bins with \rpl$\leq4$\rearth\ and an orbital period $\leq$200~days into these to obtain the values for the new grid. For the 200-400~day period bin we assume the values were identical to the 100-200 day bin. For all bins above 4\rearth\ we use the values for K dwarf stars but reduced by a factor of 0.5 to account for the generally lower occurrence rates of giant planets around M dwarfs \citep{2021CARMENES, 2023Bryant, 2023Gan}.

To account for the fact that some bins in the grid from \cite{2015Dressing} are unconstrained and contain only an upper bound on the occurrence rate we manually set the equivalent grid cells in our new grid to be unconstrained as well. For all such grid cells, we set the median and lower bound values to zero and the upper bound value to the interpolated median value added to the interpolated upper bound value. We consider the new grid cells between 0.78-1.56~days and 2-4\rearth, 6.25-25~days and 1-1.41 \rearth,  50-200~days and 0.5-1\rearth\ and 2.83-4\rearth, to be equivalent to the old grid cells between 0.5-1.7~days and 2-4\rearth, 5.5-18.2~days and 1.0-1.5 \rearth, 60.3-200~days and 0.5-1\rearth and 3.5-4\rearth respectively for the purposes of considering them unconstrained. Furthermore we also consider all the grid cells between 200-400 days and <4\rearth\ to be unconstrained as these are extrapolated and not interpolated from the measured \kepler\ rates by \cite{2015Dressing} and such planets are shown to be very rarely detected by \tess\ if at all \citep[NASA Exoplanet Archive;][]{2013Akeson}.

\begin{figure}
    \centering
    \includegraphics[width=\columnwidth]{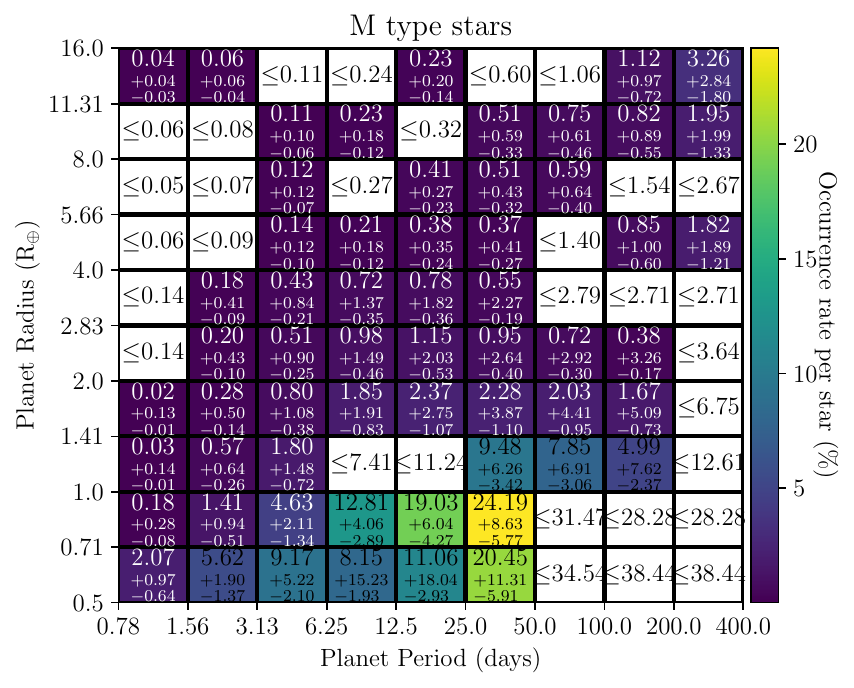}
    \caption[Rebinned M dwarf planetary occurrence rates]{Percentage occurrence rates of planets around M dwarfs.  Values for R$<$4\rearth\ taken from \protect\cite{2015Dressing}, rebinned to our standardised grid.  Values for R$>$4\rearth\ from the a scaled K dwarf occurrence rate from \protect\cite{2020Kunimoto}.}
    \label{fig: M rates}
\end{figure}

Our resulting M-dwarf occurrence rate grid is set out in Figure~\ref{fig: M rates}.

The \cite{2015Dressing} occurrence rates were calculated prior to Gaia, and therefore the stellar parameters of candidates are not as robust as post-Gaia studies such as \cite{2019Hsu, 2020Kunimoto}. Recently; \cite{2023Bergsten} have updated the \cite{2015Dressing} occurrence rates for small radius planets (\rpl$<$2\,\rearth). These small radii planets are unlikely to be detectable in the \tess\ data, so this update will not significantly impact the results we present in this paper. However, we plan to include these updated occurrence rates in future \tiara\ simulations.

\subsubsection{Use as input priority metrics}
As set out in Section~\ref{subsubsection:planet_params} we also use occurrence rate grids as a weighted distribution to select periods and radii for planets we inject into the simulation. This requires us to set a median value for unconstrained grid cells that possess only an upper bound. For the A, F, G and K dwarf grids we perform a linear extrapolation of each grid row and column and then take the mean of these as the input grid. For the M dwarf grid we simply take the upper bound of unconstrained grid cells as the median value.

\subsection{Yield Estimates}
\label{subsection:yields}
To calculate our final yield estimates we multiply the occurrence rates by the detection efficiencies for each spectral type, and then multiply those rates and their uncertainties by the number of each type of star in the Year 1 and Year 3 SPOC FFI sample. This gives us a grid of expected yields for each period-radius bin with uncertainties combined from those of the detection sensitivities and occurrence rates. The error in occurrence rate is considerably larger and so dominates the overall uncertainty in predicted yield.

\section{Results}
\label{section:results}
\subsection{\tess\ Sensitivity}
\label{subsection: Sensitivity maps}

\begin{figure}
    \centering
    \includegraphics[width=\columnwidth]{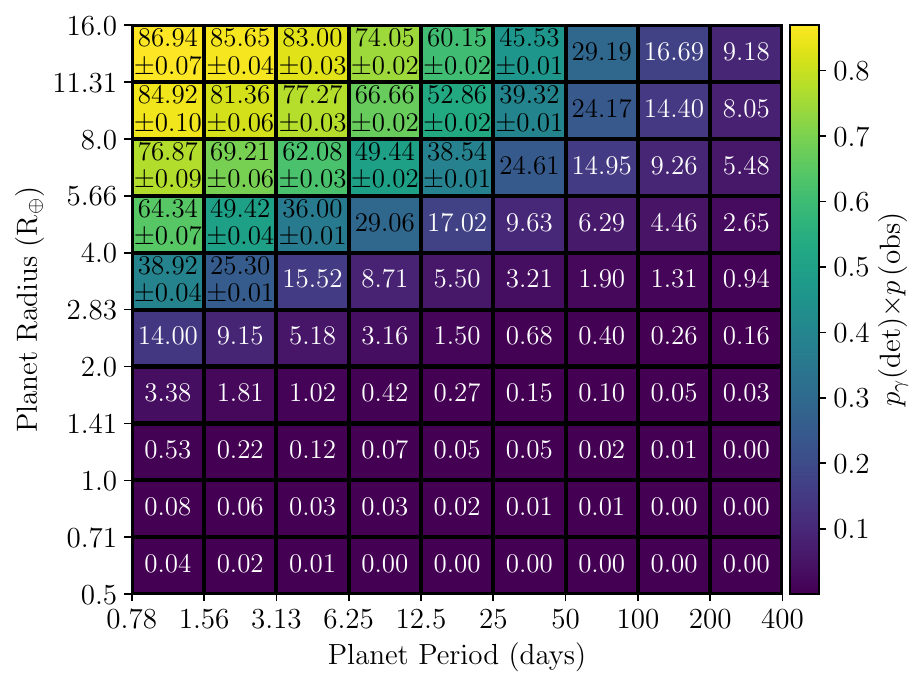}
    \caption[\tess\ Y1+3 sensitivity map]{Sensitivity map showing the probability of a transiting planet in each period/radius bin being observed and detected around an AFGKM dwarf star in the \tess\ Year 1 and Year 3 SPOC FFI lightcurves. The percentage sensitivity values are shown in each grid cell. Where no uncertainties are given the uncertainty is 0\% to two decimal places.}
    \label{fig: sensitivity_map_all}
\end{figure}

The \tiara\ pipeline creates binned sensitivity maps on the same grid as the occurrence rates from \cite{2020Kunimoto}. These show the probability that a transiting exoplanet of a given radius and orbital period is observed and detected by \tess\ in the Year 1 and Year 3 SPOC FFI lightcurves.  We compute separate sensitivity maps for each of A,F,G,K, and M dwarfs, and these are set out in Appendix~\ref{section:appendix_maps} Figure~\ref{fig:sensitivity_types}.  We also plot the average sensitivity for AFGKM dwarfs in the SPOC FFI sample in  Figure~\ref{fig: sensitivity_map_all}. The uncertainties in sensitivity are very small, this is largely on account of the large numbers of signals we inject in the simulation, which allows for such precise calculation.

Naturally, there is a trend of higher sensitivity towards larger radii planets as these planets produce deeper transits and thus a higher SNR signal.  We find that \tess\ should detect over 80\% of very short period (P$<$6.25\,day), transiting giant planets (R$>$8\rearth).  This probability drops below 50\% for small transiting planets (R$<$4\rearth), and is below 1\% for Earth-radii transiting planets.  

Additionally, we are more sensitive to shorter period planets due to the greater number of observed transits in any given monitoring duration. This both produces a higher SNR value when these transits are summed and allows for greater confidence in the detection as accounted for by the gamma function we use to calculate probability of detection (see Section~\ref{subsubsection: p_det}).  For giant planets (R$>$8\rearth), we move from approximately 80\% completeness for short periods (P$<$6.25\,day) down to less than 10\% completeness for periods between 200 and 400\,days. 

As the \tess\ mission continues, these sensitivity maps will evolve as a function of the window function for each star, and the sensitivity will improve over the entire grid. \tiara\ will be able to re-calculate these sensitivity maps using the timestamps from TESS lightcurves as the extended \tess\ mission continues to gather data.  For the southern ecliptic hemisphere, this will be the update provided by the Year 5 lightcurves (Sectors 61-69).  It is also possible to simulate timestamps for future sectors based on our knowledge of previous sectors with some assumptions needed for distribution of the future data gaps and FFI cadence.

\subsection{Predicted Yield}
\label{subsection: yield}

\begin{figure}
    \centering
    \includegraphics[width=\columnwidth]{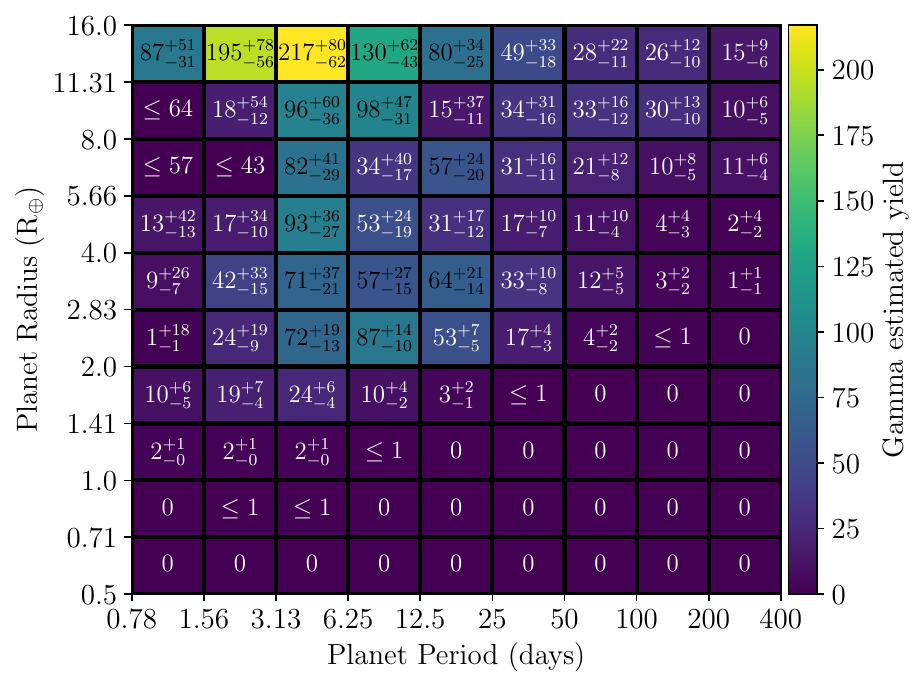}
    \caption[\tess\ Y1+3 yield]{Predicted exoplanet yield from Year 1 and 3 SPOC FFI lightcurves of AFGKM type stars. Note that cells where the median value is 1 have been displayed as $\leq1$ to account for the inherent uncertainty in a predicted yield of 1.}
    \label{fig: AFGKM_yield_grid}
\end{figure}

\begin{figure*}
    \begin{subfigure}{\columnwidth}
        \centering
        \includegraphics[width=\columnwidth]{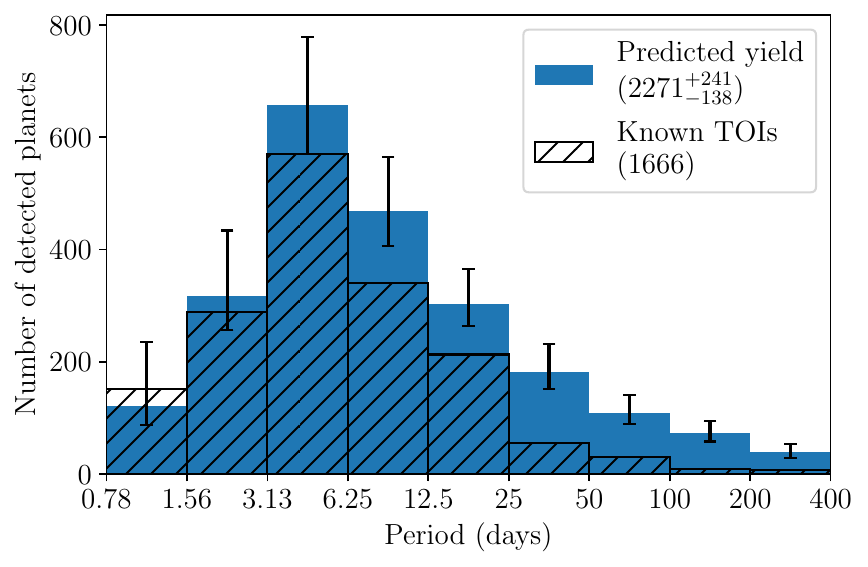}
        \caption{Distribution of orbital periods of predicted yields}
        \label{fig: yield_period}
    \end{subfigure}
    \begin{subfigure}{\columnwidth}
        \centering
        \includegraphics[width=\columnwidth]{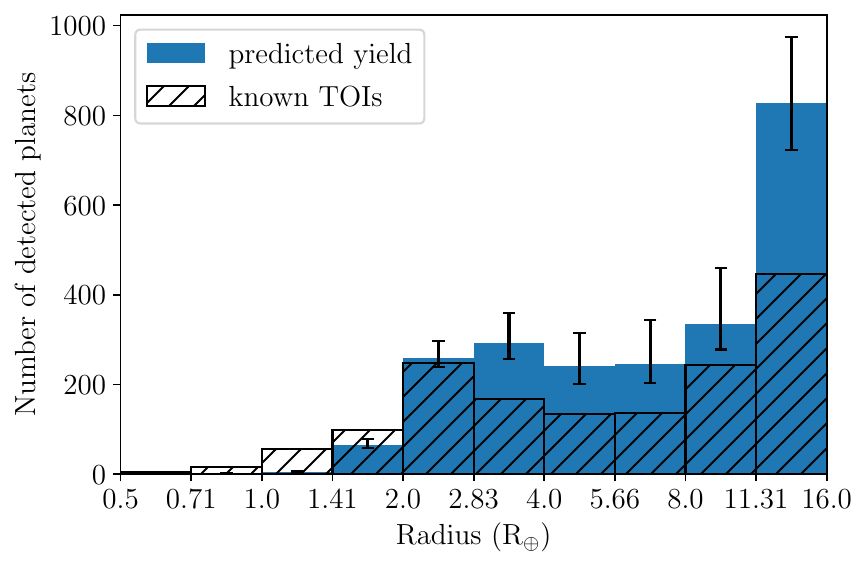}
        \caption{Distribution of planetary radius of predicted yields}
        \label{fig: yield_radius}
    \end{subfigure}
    \begin{subfigure}{\columnwidth}
        \centering
        \includegraphics[width=\columnwidth]{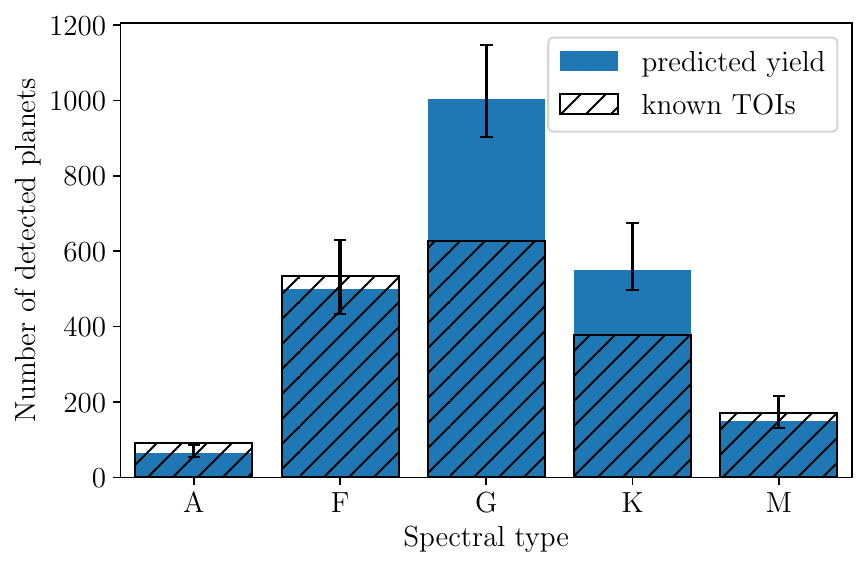}
        \caption{Expected yield by spectral type}
        \label{fig: yield_type}
    \end{subfigure}
    \begin{subfigure}{\columnwidth}
        \centering
        \includegraphics[width=\columnwidth]{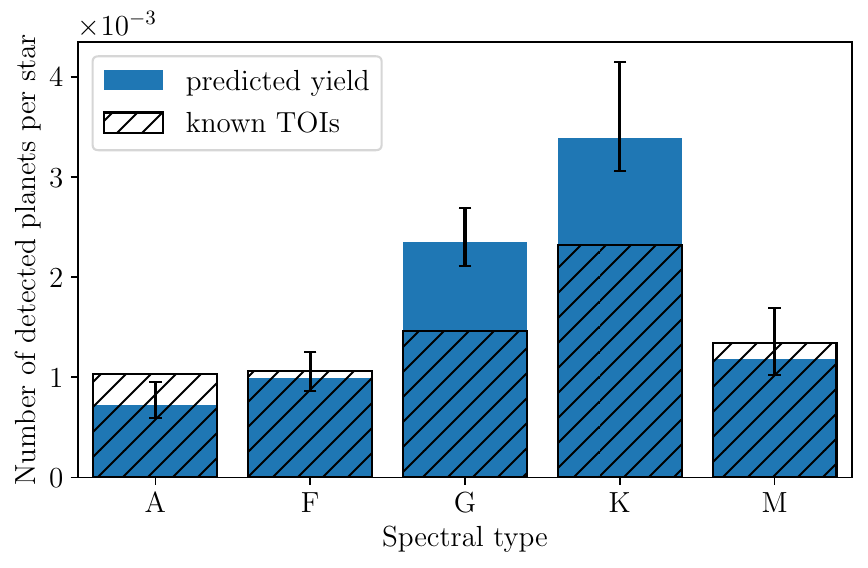}
        \caption{Expected yield per star by spectral type}
        \label{fig: yield_type_norm}
    \end{subfigure}
    \caption[Total yield histograms]{Predicted \tiara\ transiting exoplanet yields from the \tess\ Year 1 and Year 3 SPOC FFI lightcurves (solid blue bars).  Also displayed are actual \tess\ discoveries (transparent black outlined bars) calculated using TOI catalogue downloaded from the NASA exoplanet archive \citep{2013Akeson} on 2023-06-15 (excluding flagged false positives) and matched to TIC IDs of southern ecliptic SPOC FFI sample.}
    \label{fig: yield_summary}
\end{figure*}

We present the \tiara\ predicted yield of transiting exoplanet discoveries from \tess\ Year 1 and Year 3 SPOC FFI lightcurves in Figure~\ref{fig: AFGKM_yield_grid}.  This grid is the summation of all spectral types, however we also present the yields broken down by spectral type in Appendix~\ref{section:appendix_yield} Figure~\ref{fig:yield_types}.

Overall we predict a yield of $2271^{+241}_{-138}$ exoplanets detected around AFGKM dwarf host stars. We set out the yield distributions of orbital period, planet radius, and host spectral type in Figure~\ref{fig: yield_summary}.  We find the discoveries should peak at orbital periods between 3.13 and 6.25\,days.  However there are a significant number of longer period planets in our predicted yield, with $403^{+64}_{-38}$ planets with orbital periods greater than 25\,days and $113^{+23}_{-17}$ will have orbital periods greater than 100\,days.  Interestingly the distribution of planet radii is quite flat, with the exception of giant planets (R$>$11.31\rearth), which are twice as numerous as other radius bins.  Most discoveries are predicted to be around G dwarf stars, although K dwarfs provide the most number of discoveries per star monitored.

To compare our predicted yield to the actual discoveries, we crossmatch the TOI catalogue as of 2023-06-15 with the SPOC FFI target lists for Year 1 and 3 to find transiting exoplanet discoveries from our sample. The distributions of these TOIs and our predicted yields are shown in Figure~\ref{fig: yield_summary}.

\subsection{Monotransits}
\label{subsection: mono yield}

\begin{figure*}
    \begin{subfigure}{\columnwidth}
        \centering
        \includegraphics[width=\columnwidth]{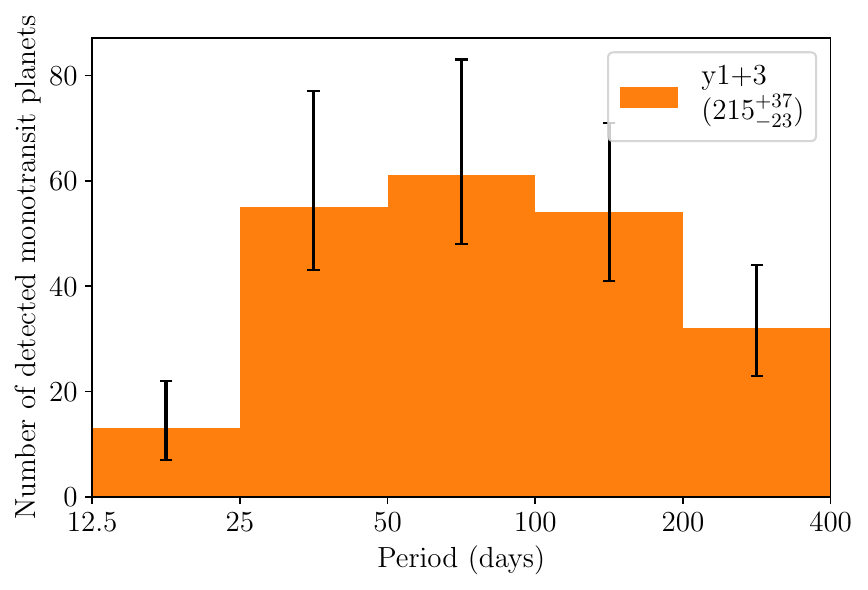}
        \caption{Period distribution of predicted monotransit detections}
        \label{fig: yield_period_mono}
    \end{subfigure}
    \begin{subfigure}{\columnwidth}
        \centering
        \includegraphics[width=\columnwidth]{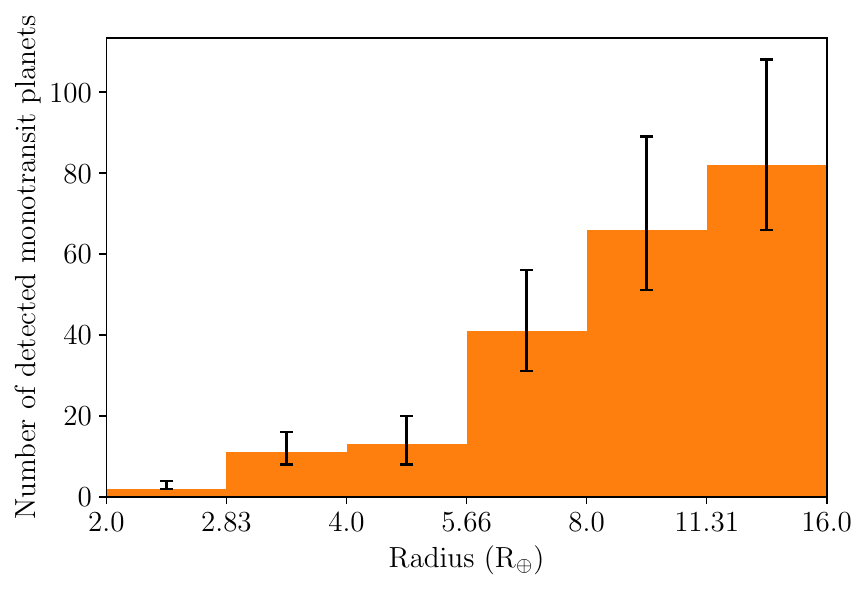}
        \caption{Radius distribution of predicted monotransit detections}
        \label{fig: yield_radius_mono}
    \end{subfigure}
    \caption[Monotransit yield histograms]{Predicted yields of monotransits, shown as binned distributions in orbital period (a) and radius (b).}
    \label{fig: yield_mono}
\end{figure*}

We predict that $215^{+37}_{-23}$ of the transiting planet discoveries will only have one transit event in the \tess\ Year 1 or Year 3 data - i.e. are "monotransits".  This makes up a relatively small fraction (9\%) of the total predicted yield, but is a significiant fraction of the long-period detections.  $202^{+36}_{-22}$ (50\%) of the planets with $P>25$\,days are monotransits, while $86^{+20}_{-14}$ (76\%) of the planets with $P>100$\,days are monotransits. Figure~\ref{fig: yield_mono} shows the the distributions of orbital period and planetary radius for the predicted monotransit yield.  As expected there are very few monotransits with orbital periods less than the sector length of \tess\ (27\,days).  However beyond that the distribution of orbital period is remarkably flat out to our final bin of 400\,days.  The distribution of planetary radii for the monotransits are much more heavily skewed towards larger radii planets than typical multi-transit detections. 73\% of all detections are for planets with $Rp>4$\rearth, while this fraction is 94\% of just the monotransit detections.

\begin{figure}
    \centering
    \includegraphics[width=\columnwidth]{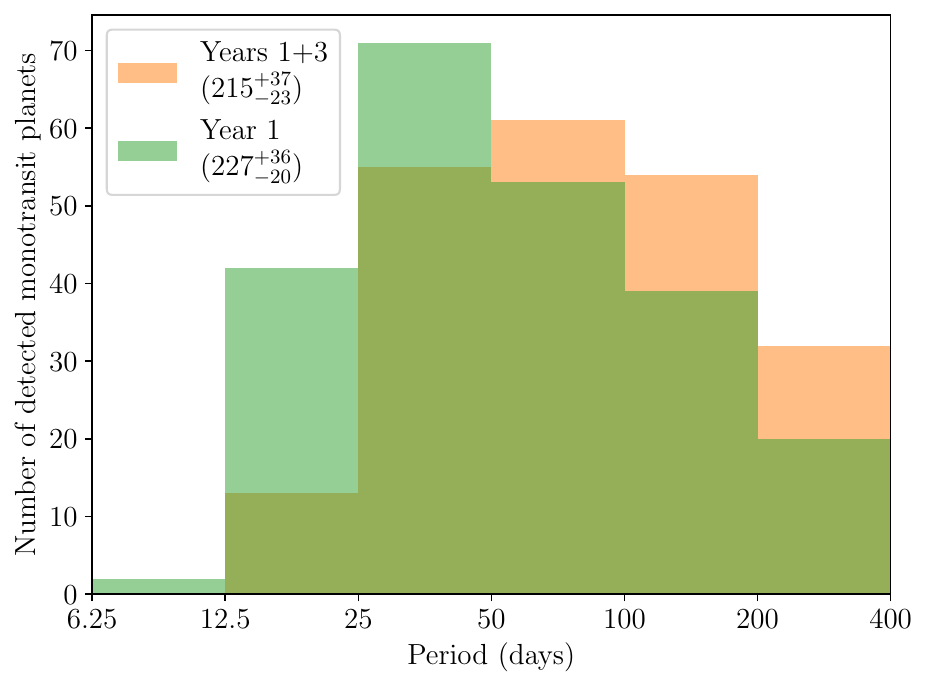}
    \caption[Monotransit period evolution]{Comparison of the distribution in period of the yields of monotransits from Year 1 alone (green) compared to Year 1 and Year 3 combined (orange).}
    \label{fig:monos_evolution}
\end{figure}

In addition to the monotransit yield for Years 1 and 3 together, we also estimated the yield for Year 1 alone in order to investigate the effect of reobserving the field on the monotransit yield. The results are shown in Figure~\ref{fig:monos_evolution}. We find that the number of monotransits in Year 1 alone is $227^{+36}_{-20}$, compared to the Year 1 and 3 combined yield of to $215^{+37}_{-23}$.  Assuming that all of the Year 1 monotransits that have an additional Year 3 transit observed become duotransits (see Section~\ref{subsection: duo yields}), then this leaves $114^{+42}_{-24}$ detected monotransits observed in Year 1 and not Year 3 and $101^{+55}_{-33}$ observed in Year 3 and not Year 1.  The distribution of orbital periods changes to favour longer periods with 40\% of Year 1 and Year 3 monotransits having P$>$100\,days compared to 26\% of Year 1 only monotransits. The distribution of monotransit detections in period also becomes significantly flatter at longer-periods when Year 3 is added to Year 1.

\subsection{Duotransits}
\label{subsection: duo yields}

\begin{figure*}
    \begin{subfigure}{\columnwidth}
        \centering
        \includegraphics[width=\columnwidth]{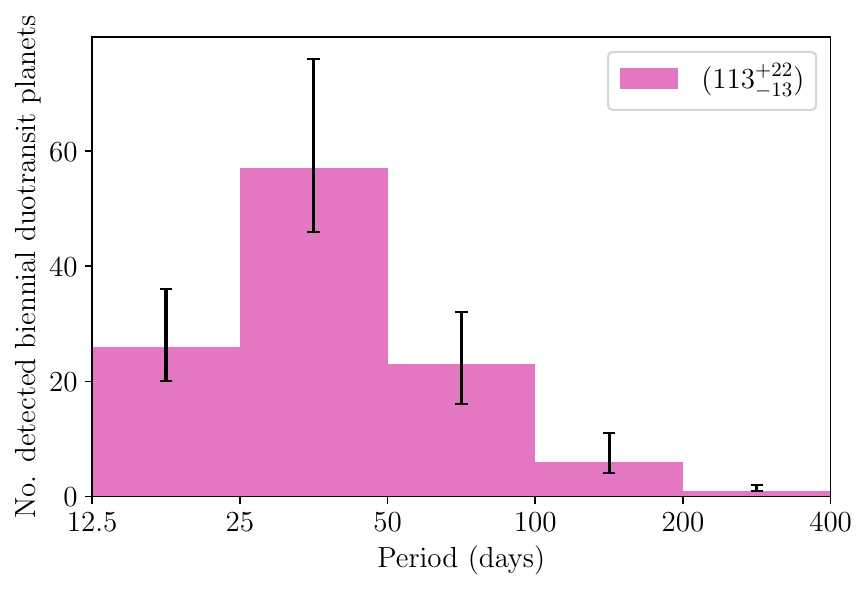}
        \caption{Period distribution of predicted biennial duotransit yield}
        \label{fig: yield_period_true_duo}
    \end{subfigure}
        \centering
        \begin{subfigure}{\columnwidth}
        \includegraphics[width=\columnwidth]{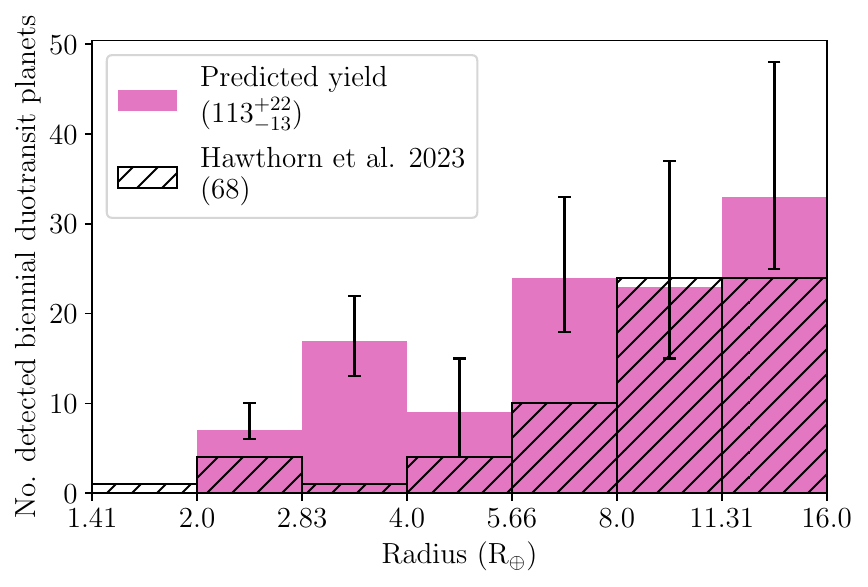}
        \caption{Radius distribution of predicted biennial duotransit yield}
        \label{fig: yield_radius_true_duo}
    \end{subfigure}
    \caption[Biennial duotransit yield histogram]{Predicted yields of biennial duotransit planets detected from one transit in Year 1 and another in Year 3. Shown in distributions of period (a) and radius (b). Plot (b) also shows a comparison with the radii of biennial duotransits from \citet{DuosHawthorn} excluding candidates over 16\rearth\ which we do not simulate.}
    \label{fig:duos_hists}
\end{figure*}

As \tess\ reobserved the southern ecliptic hemisphere in Year 3 of the mission, many Year 1 monotransit signals will have an additional transit observed in Year 3, and become "biennial duotransits".  We define these duotransits as "biennial" to distinguish them from other duotransits in which the two transit events happen in the same year of \tess\ data. . We predict a total of $170^{+29}_{-18}$ duotransits in total with $113^{+22}_{-13}$ of these being biennial duotransits.

The distribution of biennial duotransits is shown in Figure~\ref{fig:duos_hists}. Compared to monotransit detections (see Section~\ref{subsection: mono yield}) we find that the sample of planets detected from biennial duotransits show a flatter distribution in radius, with 21\% of biennial duotransit detections coming from planets below 4\rearth\ compared to 6\% of monotransits, although radii $>$11.31\rearth\ are still the largest bin of detections. We find the period of expected biennial duotransits peaks between 25 and 50\,days and falls off more rapidly than monotransits for longer periods, with 6\% of biennial duotransits coming from periods $>$100\,days compared to 40\% of monotransits. 

\subsection{Breakdown by Spectral Type}
\label{subsection: spectral type yields}
As shown in Figure~\ref{fig: yield_type}, we find G dwarfs should be host to the most detections $(1005^{+143}_{-103})$, followed by K $(551^{+123}_{-54})$ and F $(500^{+130}_{-67})$ dwarfs in quick succession with a reasonable number from M dwarfs $(151^{+66}_{-20})$ and a very small number from A dwarfs $(64^{+21}_{-11})$. This distribution is largely due to the numbers of each spectral type represented in the SPOC FFI sample with G types being the most numerous and thus host to the most detected planets. 

Figure~\ref{fig: yield_type_norm} shows the yield from each spectral type divided by the number of stars of that spectral type in the SPOC sample. From this it becomes apparent that K dwarfs are the most numerous hosts of detections per star with a rate of $(3.4^{+0.8}_{-0.3})\times10^{-3}$ transiting planets per star observed. G dwarfs are the next most efficient sources of detections with $(2.4^{+0.3}_{-0.2})\times10^{-3}$ predicted per star, followed by M with $(1.2^{+0.5}_{-0.2})\times10^{-3}$, F with $({0.99}^{+0.26}_{-0.13})\times10^{-3}$ and A with $(0.72^{+0.24}_{0.12})\times10^{-3}$.

K dwarfs sit at the best crossover between occurrence rates and \tess' detection efficiency, and thus have the highest rate of detections per star.  Despite M dwarfs having higher planetary occurrence rates than AFGK stars \citep{2015Dressing} and their small size causing deeper transits due to a higher planet-star radius-ratio, they display a relatively low rate of detections per star. M dwarfs host many small radius planets compared with AFGK dwarfs, but fewer giant planets. These small planets produce low SNR signals that result in low detection probabilities for M dwarf host stars. This results in fewer detections per star for M dwarfs.

\begin{figure}
    \centering
    \includegraphics[width=\columnwidth]{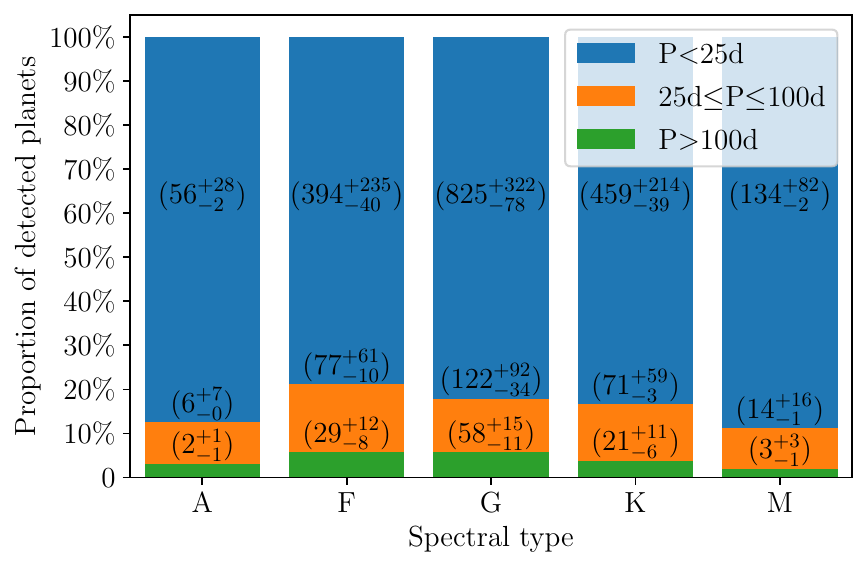}
    \caption[Long period yields by spectral type]{Comparison of the proportions of predicted long-period planet yield by spectral type. Planets with periods less than 25 days are shown in blue, periods between 25 and 100 days are shown in orange and periods longer than 100 days are shown in green.  The predicted yield numbers for each spectral type and period range are printed on the plot, along with the associated uncertainties.}
    \label{fig: yield_type_period}
\end{figure}

Figure~\ref{fig: yield_type_period} shows the proportion of different period ranges in the predicted yield by spectral type. We find a reasonably flat trend in the distribution of long-period discoveries for each spectral type with a possible minor trend towards shorter-period discoveries being favoured at later spectral types, although increasingly small numbers create uncertainty in such a trend.

\subsection{Comparison on Detection Probability Functions}
\label{subsection:DPF}

\begin{figure*}
    \begin{subfigure}{\columnwidth}
        \centering
        \includegraphics[width=\columnwidth]{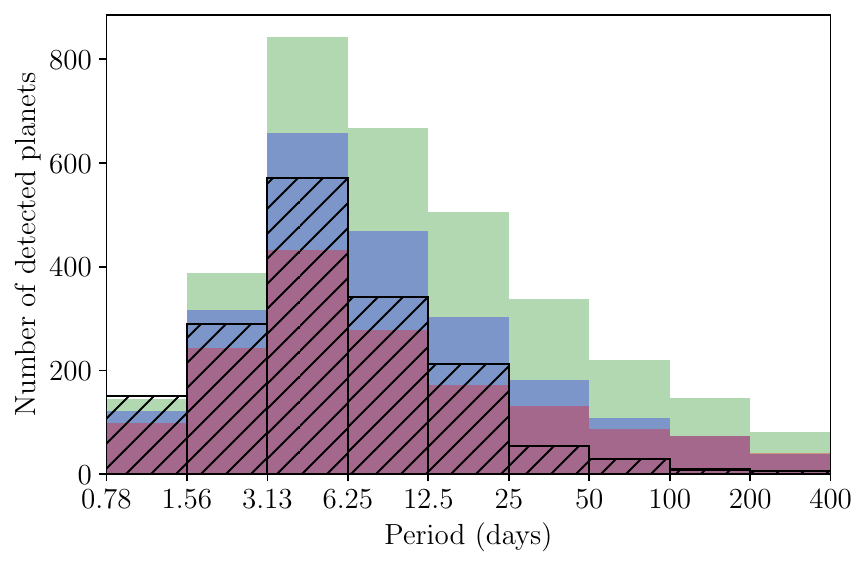}
        \caption{Period distribution of predicted yields for three detection criteria}
        \label{fig:methodcompare_period}
    \end{subfigure}
    \begin{subfigure}{\columnwidth}
        \centering
        \includegraphics[width=\columnwidth]{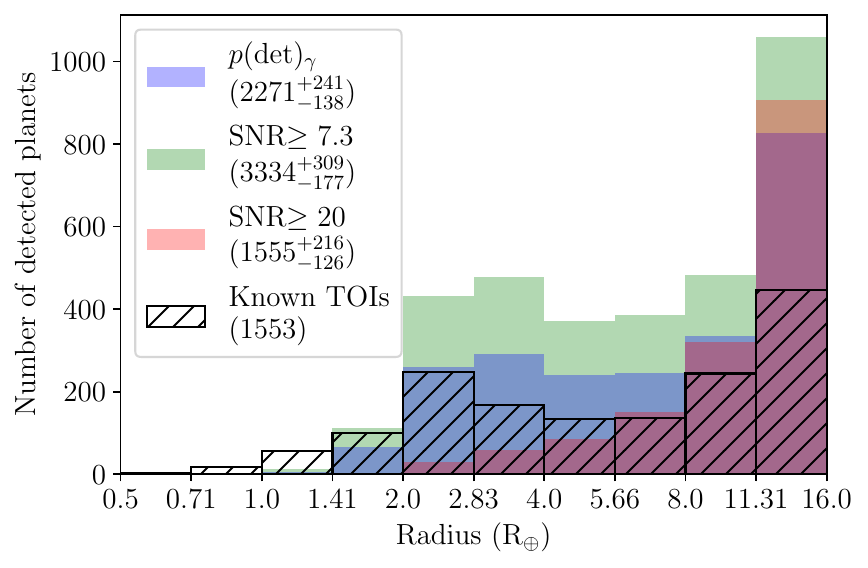}
        \caption{Radius distribution of predicted yields for three detection criteria}
        \label{fig:methodcompare_radius}
    \end{subfigure}
    \caption[Yields for different detection criteria]{Comparison of the yields predicted using three different detection criteria, a gamma cumulative distribution (blue), and signal to noise thresholds of 7.3 (green) and 20 (red). Distributions are shown in period (a) and radius (b).}
    \label{fig:methodcompare}
\end{figure*}

As well as the gamma function described in Section~\ref{subsection:threshold} we also calculate yields using simple signal to noise threshholds of 7.3 and 20. Figure~\ref{fig:methodcompare} shows a comparison of the results from these three approaches.

The predicted total yield using a $\text{SNR}\geq20$ threshold of $1555^{+216}_{-126}$ shows a strong agreement with the total number of TOIs from the Year 1 and 3 SPOC FFI discoveries (1666). However Figure~\ref{fig:methodcompare} demonstrates that the predicted distribution in periods and radii from this criterion does not match the proportions of the TOI sample as well as other methods. We see a significant under-prediction of planet yields for radii below 5.66\rearth\ which increases in disagreement for decreasing radii from this method as shown in Figure~\ref{fig:methodcompare_radius}. We also find that the yields for short period planets ($P<25$\,days) are under-predicted by this method, with the largest deficit being between 3.13 and 6.25 days

Using a smaller SNR threshold of 7.3 as many yield studies have done previously \citep[e.g;][]{2015Sullivan, 2017Bouma, 2018Barclay, 2018Cooke, 2019Cooke, 2019Villanueva}, we find a significantly greater yield prediction than either the threshold of SNR$\geq20$ or the gamma function. The predicted yield of planets with SNR$\geq7.3$ ($3334^{+309}_{-177}$) is more than double the size of the TOI discoveries and the yield predicted by SNR$\geq20$ and around 1.5 times greater than that predicted by the gamma function. A large fraction of this additional predicted yield comes from small planets and longer period planets as shown in Figure~\ref{fig:methodcompare}.

\section{Discussion}
\label{section:discussion}
\subsection{Comparison to actual \tess\ detections}
\label{subsection:tess-yield-comparison}
Comparison to the population of \tess\ transiting planet discoveries via the TOI catalogue is a relatively straightforward and robust check for the \tiara\ yield results. We cross-match the TIC IDs of the TOI catalogue with stars we simulated from the SPOC FFI sample to compare against our predicted yields as shown in Figure~\ref{fig: yield_summary}.

While we use the SPOC FFI lightcurves as a homogeneous sample it must be noted that the TOI sample is not generated homogeneously. While we limit the TOI catalogue to just those which have SPOC FFI lightcurves which we include in our simulation, many of these will still have been discovered in lightcurves produced by other pipelines such as the SPOC 2 minute postage stamp or Quick look lightcurves \citep[QLP;][]{2020Huang}. These lightcurves may have different duty cycles, noise levels and systematics to the SPOC FFI sample we use. Different quality flags in these pipelines may also affect the location and duration of data gaps between data products. Additionally the automated TOI search carried out is different between pipelines, which means TOIs from different pipelines are affected by different biases. It is for this reason that we stress that our comparison is not specifically an attempt to assess the completeness of the SPOC FFI pipeline TOI search.
 
\begin{figure}
    \centering
    \includegraphics[width=\columnwidth]{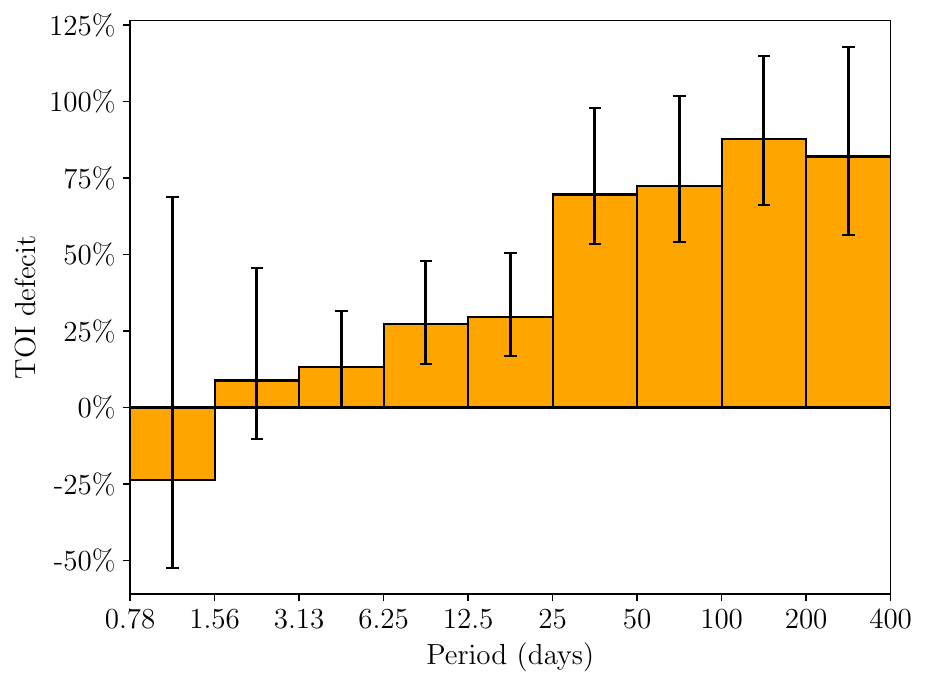}
    \caption[Long-period deficit]{Percentage deficit between the discovered TOIs and our predicted yield for the \tess\ SPOC FFI lightcurves from Year 1 and Year 3.}
    \label{fig:yield_period_residuals}
\end{figure}

Our total predicted yield of $2271^{+241}_{-138}$ is $\sim4.4\sigma$ greater than the total number of TOIs (1666) as of 2023-06-15. However, for orbital periods less than 6.25 days the \tiara\ predictions agree with he discoveries to within uncertainties.  For periods between 6.25 and 25 days the \tiara\ predictions are not more than $2\sigma$ above the discoveries. However for planets with periods greater than 25\,days the deficit between \tiara\ predictions and TOI discoveries becomes increasingly significant as shown in Figure~\ref{fig:yield_period_residuals}.

The SPOC Transiting Planet Search \citep[TPS;][]{2016Jenkins, 2021Guerrero} relies on phase folding data to search for periodic signals \citep{2021Guerrero}. This becomes increasingly difficult for longer-periods as such planets may have transits in different \tess\ sectors or only one transit which is impossible to phase fold. This suggests that the difference between our predicted yield and the TOI sample may be due to \tess\ discoveries being currently incomplete for longer-period planets. Since the majority of candidates are given TOI status by the TPS \citep{2021Guerrero}, many monotransit and biennial duotransit events in the SPOC FFI lightcurves may have yet to be detected.

It is difficult to uniformly simulate the planet detection process that makes up the TOI catalogue, as the TOI catalogue is not built up from a single uniform search of the SPOC FFI lightcurves. Instead the SPOC Transiting Planet Search only runs on selected postage stamp lightcurves (around 20,000 per sector) while the rest of the SPOC FFI targets are searched by Quick Look Pipeline \citep[QLP;][]{2020Huang}. However, the QLP operates at a hard magnitude cutoff of T=13.5, and some of the SPOC FFI targets are fainter than this cut-off. Therefore some of the lightcurves in the SPOC FFI sample have not been searched in a systematic manner.  Additionally, third party groups will have their own methods of searching lightcurves for candidates which can lead to additional TOIs being created via the Community TOI process \citep[CTOI;][]{2021Guerrero}. These subtleties could contribute to a small discrepancy between our predictions and the actual TOI catalogue, although significant differences are likely still due to incompleteness as described previously.

We can gain an indication of this by comparing our predicted yield with and without monotransits to the TOI sample. When we remove monotransits from our predicted yield we find $2056^{+245}_{-140}$ detected exoplanets and excluding biennial duotransits we further reduce this to $1943^{+246}_{-141}$. This brings our yield into closer alignment with the TOI detections \citep[Taken from NASA exoplanet archive;][]{2013Akeson} from $4.4\sigma$ to $2.8\sigma$ and $2.0\sigma$ respectively.

This could also explain the seeming overabundance of giant planets as seen in Figure~\ref{fig: yield_radius} as long-period monotransit and duotransit detections are particularly biased towards larger radii (see Figures~\ref{fig: yield_radius_mono}~and~\ref{fig: yield_radius_true_duo}), and hence the undiscovered long-period planets make up the majority of the undiscovered giant planets as well. Additionally it could explain the overestimation of planets detected around G and K dwarfs shown in Figures~\ref{fig: yield_type}~and~\ref{fig: yield_type_norm}. As these stars make up the largest numbers of stars in the SPOC FFI sample and have the highest detection rate per star, it it likely that most of the long-period and other planets not yet found would be around these.

\subsection{Comparison to Biennial Duotransit search}
\label{subsubsection:faith-comparison}
We are able to compare our predicted yield of biennial duotransit detections with a real search of the Year 1 and Year 3 SPOC FFI lightcurves for biennial duotransits set out in  \citet{DuosHawthorn}. Figure~\ref{fig: yield_radius_true_duo} shows the distribution of the \tiara\ predicted yields and the sample from \citet{DuosHawthorn} in radius.  The periods of the majority of candidates in \citet{DuosHawthorn} are currently unknown and so cannot be compared directly to our prediction. Our predicted yield of biennial duotransits is $113^{+22}_{-13}$, which is $3.5\sigma$ greater than the 68 candidates in the real sample. The numbers are in close agreement for radii between 8.0 and 11.31\rearth\, and near agreement for 11.31 to 16\rearth.  However for smaller radii planets (R$<4$\rearth), \tiara\ predicts many more biennial duotransits than than are detected in \citet{DuosHawthorn}, especially between 2.83 and 4 \rearth. This is to be expected as \citet{DuosHawthorn} only consider high SNR biennial duotransits, which will be produced by larger planets. We note that \citet{DuosHawthorn} also have a significant number of candidates with radii larger than 16\rearth, however we do not consider such large radius objects in our \tiara\ simulation. 

We also note the strong agreement between the predicted number of biennial duotransits with SNR$\geq20$ of $86^{+25}_{-13}$ and the discovery of 68 biennial duotransits from \citet{DuosHawthorn}. The effect of using different probability of detection functions is discussed further in Section~\ref{subsection:p_det}.

\subsection{Comparison to previously predicted yield}
\label{subsection:previous-yield-predictions}
In addition to comparing our results to the TOI catalogue (see Section~\ref{subsection:tess-yield-comparison}) we can also check the validity of our results by comparing them to previous attempts at predicting yield. This is especially useful as our methods include monotransits, which are not included in studies such as \citet{2022Kunimoto}. 

\citet{2022Kunimoto} use a considerably larger population of stars.  We have simulated the yield from the $\sim1.3$\,million stars in the Year 1 and Year 3 SPOC FFI sample. \cite{2022Kunimoto} use 4 million stars from the \tess\ Candidate Target List \citep[CTL;][]{2019Stassun}. Therefore we normalise each set of results by the number of stars in each sample and compare in terms of number of planets predicted per star (similarly to in Figure~\ref{fig: yield_type_norm}) for ease of comparison. However, it should be noted that the selection criteria for each stellar sample, especially for SPOC stars \citep{2020Caldwell}, introduces biases in the average detectability for each stars, so this is somewhat of a crude comparison.

\cite{2022Kunimoto} predict a total of $2532\pm189$ exoplanets found around 4066063 stars in Year 1 and $1748\pm103$ around 4021948 stars in Year 3. This gives a total of $(1.06\pm0.05)\times10^{-3}$ detections per star. This is 6.5 $\sigma$ below our estimate of $(1.55^{+0.18}_{-0.11})\times10^{-3}$ per star excluding monotransits. It is to be expected that our yield prediction will be greater than  \citet{2022Kunimoto} since we use the SPOC FFI sample, which has been explicitly created to target stars that are most likely to have detectable transits with \tess\ \citet{2020Caldwell}.

\subsection{Eccentricity distribution}
\label{subsection:ecc-discussion}
As in \citet{2018Cooke,2019Cooke} we use a beta distribution to assign eccentricity to planets we inject into our simulation and use values for $\alpha$ and $\beta$ taken from \cite{2015VanEylen}. However, this study was focused on super-Earths and Mini-Neptunes from \kepler and not longer-period giant planets which are the focus of this paper. Warm-Jupiters have been found to often show much higher orbital eccentricities than smaller and/or closer-in planets \citep{2021Dong} and therefore we are likely underestimating the eccentricity of the Warm-Jupiter population in our simulation. However eccentric long-period planets are more likely to transit at periastron when closer to their host stars and hence this means we are likely \textit{underestimating} the detection sensitivity of \tess\ towards such planets.

\subsection{Occurrence Rates}
\label{subsection:occurrence}
The \tiara\ yield predictions rely on the occurrence rates from \cite{2020Kunimoto} for AFGK stars and a modified  occurrence rate from \cite{2015Dressing} for M dwarfs.

The occurrence rates we use from \cite{2020Kunimoto} do not include any planets larger than 16\rearth\ although larger planets are known to exist and are likely to be detectable by \tess\ even if the occurrence rates from \kepler\ are known to be low \citep{2013Fressin, 2019Hsu, 2020Kunimoto}. Future yield estimate studies could benefit from studies into the occurrence rate of planets larger than 16\rearth which could be carried out on samples of planets from \tess.

Due to the lack of occurrence rate studies on A and earlier spectral-type stars, we assume the occurrence rates of planets around A and earlier type stars are equal to that of F type stars (see Section~\ref{subsection:occurance_rates}). In reality these may be different as the higher mass and temperature of A type stars are expected to have significant effects on planet formation and thus the occurrence rate \citep{2010Johnson}. Future studies could benefit from \tess\ derived occurrence rates of A type stars \citep[e.g;][]{2019Zhou, 2023Johnson}.

For M dwarfs we rebin the values found by \cite{2015Dressing} onto the same grid used by \cite{2020Kunimoto}. While this does, of course, change the exact values for M dwarf occurrence rates, due to the large errors on the values from \cite{2015Dressing} we believe our values for occurrence rates within the original bounds of this grid (0.5\rearth$\leq$\rpl$\leq$4\rearth, 0.5~days$\leq P\leq$200~days) are robust. As discussed in Section~\ref{subsection:occurance_rates} we expect the yields of planets with periods between 200-400 days from \tess\ to be low and thus treat these bins as unconstrained with only an upper bound value.  Furthermore we estimate values for planets between 4 and 16 \rearth\ around M-dwarfs using the values found for K dwarf stars by \cite{2020Kunimoto} but reduced by a factor of one half to account for the expected lower occurrence rate of giant planets around M dwarfs \citep{2021CARMENES, 2023Bryant, 2023Gan}.  To improve the accuracy of future yield estimations, it would be useful to constrain giant planet occurrence rates around M dwarfs as discussed in \citet{2023Bryant, 2023Gan} and \citet{2021CARMENES}.  It is also worth noting that by rebinning the rates for M dwarfs we increase the shortest possible period of planet we simulate from 0.5~days to 0.78~days, this means we are likely to underestimate yields in this region of period-space as \tess\ has already found  candidates \citep[NASA Exoplanet Archive;][]{2013Akeson} and 9 confirmed planets \citep{2019Vanderspek, 2022Giacalone, 2022Luque, 2023Essack, 2023Goffo} in this region.

\subsection{Transit model}
\label{subsection:trapezoid-discussion}
We use a Trapezoidal model to approximate the shape of transits in our simulation as described in Section~\ref{subsubsection:transit_timestamp}, this assumes a flat-bottomed shape to the transit and ignores the effects of limb-darkening \citep{2015Espinoza, 2016Espinoza, 2020Agol} which may slightly increase the measured depth of a transit and thus increase signal to noise. However we expect this effect to be small in magnitude and therefore did not use further computational resources to include it in our simulation. This may lead to our sensitivity values for \tess\ towards all exoplanets to be slightly conservative. 

Additionally, this model assumes the contributions to SNR from Ingress and Egress are effectively half that of the full transit. This approximation works well enough for our purposes but use of the more accurate trapezoid model put forward in \cite{2023Kipping} may improve our results. However we expect the magnitude of this effect on our estimated SNR to be negligible.

While both of these effects are small, their inclusion in future iterations of \tiara\ could improve results, and may be worth including with more computational resources.

\subsection{Probability of detection}
\label{subsection:p_det}
As set out in Section~\ref{subsection:threshold}, we use an incomplete gamma function to estimate the probability of detection in the \tiara\ simulation following \citet{2017Christiansen, 2019Hsu} and \cite{2022Kunimoto}. As discussed in \cite{2022Kunimoto}, there are some caveats associated with the use of this probability of detection function:

\begin{enumerate}
    \item \kepler\ monitored a different population of stars and had different light curve properties to \tess. While our use of real \tess\ lightcurves allows us to account for many of the unique properties of \tess\ data we still are applying a probabilty of detection function developed for \kepler\ to a very different mission. It would be useful for work similar to that of \cite{2017Christiansen} and \cite{2019Hsu} on \kepler\ DR25 to be performed on \tess\ to gain a more accurate understanding of the false positive rate and detection efficiency of \tess. 
    \item We perform a linear extrapolation for the detection efficiency of duotransit and monotransit events in \tess.  While we believe the result to be reasonable as it produces the lower detection efficiency we expect for such events it is not based on the same rigorous testing done by \cite{2017Christiansen}~and~\cite{2019Hsu} for 3 transit and greater events. Furthermore monotransits are often not found by the same algorithms that search for multitransit events and thus the probability of detection may not scale down linearly with number of transits down to the level of a monotransit. For this reason it is also worth considering a simple SNR$\geq20$ threshold for monotransits as discussed in Section~\ref{subsection:DPF}.
\end{enumerate}

Compared with \kepler, Vetting and confirmation of \tess\ candidates is easier at the same SNR due to the fact that \tess\ observes brighter stars and benefits from the \gaia\ data releases that better characterise host stars. Therefore it is likely that, with the caveats listed above, the use of the gamma function for \kepler\ on \tess\ data is likely to lead to more conservative (i.e. lower) estimates of sensitivity and therefore yield. However, quantifying the exact difference is difficult and therefore we continue to use the \kepler\ function.

Additionally, as shown in Section~\ref{subsection:DPF} the use of a gamma function, and SNR thresholds of 7.3 and 20 all recover a deficit in the number of long-period planets compared with the sample of TOI discoveries. This provides additional evidence for the potential of undiscovered long-period planets in \tess\ data.

\subsection{Use of Year 1 and Year 3 SPOC FFI Lightcurves}
\label{subsection:SPOC}
In our simulation we use Year 1 and Year 3 SPOC FFI lightcurves which gives us a stellar population of $\sim1.3$ million stars in the southern ecliptic hemisphere. This sample is selected in part due to computational limits of SPOC \citep{2020Caldwell}, and \tess\ observes many more stars at sufficient precision to detect transiting exoplanets. Thus the total planetary yield from the \tess\ mission will be higher than those from just the SPOC FFI targets as we simulate here. However, the SPOC FFI sample prioritises bright, main sequence targets with low levels of dilution from other sources to fill the maximum of 160000 targets per sector \citep{2020Caldwell}. The Candidate Target List \citep[CTL;][]{2019Stassun}, as used by \cite{2022Kunimoto}, uses a similar prioritisation metric but with a higher number of selected targets. This difference results in a smaller sample for SPOC, but with a greater proportion of targets most amenable to small planet detection. In future, it would be interesting to run \tiara\ on a larger sample of lightcurves, such as those generated by the Quick Look Pipeline \citep[QLP;][]{2020Huang}.

\section{Summary and Conclusions}
\label{section:conclusion}
We develop the \tiara\ pipeline to simulate the sensitivity of transit surveys to detecting transiting exoplanets.  We focus in this paper on the Year 1 and Year 3 SPOC FFI sample from the \tess\ mission.  In particular we are interested in the sensitivity and yield for longer-period planets that present as monotransits or biennial duotransits in the data.  Our simulations is based on the actual stars monitored by \tess\ and the real \tess\ window functions (accounting for  discontinuities in observations).  We also use the actual noise properties of the lightcurves for each star.

We find a total of $2271^{+241}_{-138}$ exoplanets should be detected around AFGKM dwarf host stars in the Year 1 and Year 3 SPOC FFI lightcurves. Of these $403^{+64}_{-38}$ will have orbital periods greater than 25 days and $113^{+23}_{-17}$ will have orbital periods greater than 100 days. We find 4.4$\sigma$ more predicted detections than the current TOI sample size of 1666. We find an increasing disparity between our predictions and actual \tess\ discoveries from the TOI catalogue at longer-periods, suggesting that the \tess\ discovery sample is incomplete at longer-periods and more long-period planets remain to be discovered in \tess\ data. 

These additional planets will require concentrated follow-up efforts to confirm as the majority of them will be initially detected as some of the $215^{+37}_{-23}$ predicted \tess\ monotransits. 50\% of planet detections with periods above 25\,days and 76\% of planet detections above 100\,days will be monotransits. Aside from monotransits, a large portion of the remaining long-period planets will be found as biennial duotransit events with one transit in Year 1 and an additional in Year 3, with $113^{+22}_{-13}$ of these discoveries detected.

The \tiara\ pipeline developed for this project can be applied to additional \tess\ data sets such as the northern ecliptic SPOC FFI lightcurves from Years 2 and 4 and the recently released Year 5 SPOC FFI lightcurves. We intend to perform follow-up simulations using these additional SPOC data in the future. The QLP lightcurves \citep{2020Huang} could also be analysed using \tiara. Reasonable assumptions of the observing strategy and current performance of \tess\ could also be used to create simulated lightcurves and make predictions of future \tess\ extended missions. With simulated data, \tiara\ can easily be used to predict the yields from future missions, in particular the upcoming PLAnetary Transits and Oscillations of stars \citep[\plato;][]{2014Rauer} mission.

\section*{Acknowledgements}

This paper includes data collected by the \TESS\ mission. Funding for the \TESS\ mission is provided by the NASA Explorer Program. Resources supporting this work were provided by the NASA High-End Computing (HEC) Program through the NASA Advanced Supercomputing (NAS) Division at Ames Research Center for the production of the SPOC data products. The TESS team shall assure that the masses of fifty (50) planets with radii less than 4\rearth\ are determined.

We acknowledge the use of public \TESS\ Alert data from pipelines at the \TESS\ Science Office and at the \TESS\ Science Processing Operations Center.

This paper includes data collected by the TESS mission that are publicly available from the Mikulski Archive for Space Telescopes (MAST).

TR and FR are supported by STFC studentships. The contributions at the University of Warwick by SG and DB have been supported by STFC through consolidated grants ST/P000495/1, ST/T000406/1 and ST/X001121/1. 

We thank Dr Chelsea X. Huang and Dr Thomas G. Wilson for their helpful comments when reviewing this work in the form of an MSc thesis. We also would like to thank the assistant editor, Dr Sean Hodges and the anonymous reviewer for their helpful comments.
\section*{Data Availability}

The \tess\ data is accessible via the MAST (Mikulski Archive for Space Telescopes) portal at \url{https://mast.stsci.edu/portal/Mashup/Clients/Mast/Portal.html}. The binned sensitivity and yield maps used to produce Figures~\ref{fig: sensitivity_map_all},\ref{fig: AFGKM_yield_grid},\ref{fig:sensitivity_types}~and~\ref{fig:yield_types} are available in a machine readable format from an open \texttt{GitHub} repository at \url{https://github.com/TobyRodel/TIaRA-TESS-Yields}.



\bibliographystyle{mnras}
\bibliography{refs} 



\appendix
\label{Appendix}
\section{Sensitivity by spectral types}
\label{section:appendix_maps}

\begin{figure*}
    \centering
    \begin{subfigure}{\columnwidth}
        \centering
        \includegraphics[width=\columnwidth]{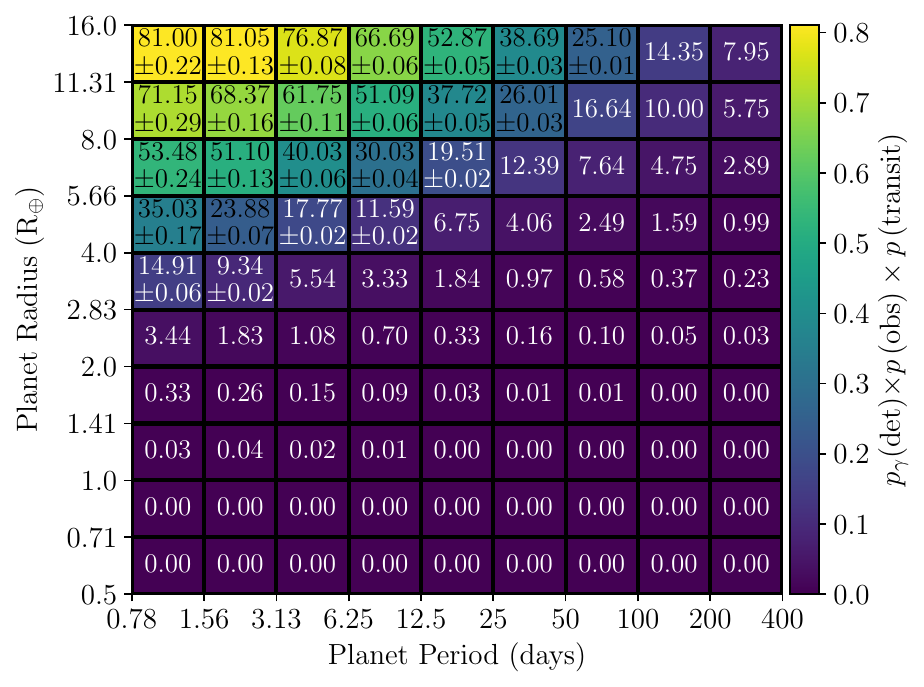}
        \caption{A type dwarf stars}
        \label{fig:sensitivity_A}
    \end{subfigure}
    \begin{subfigure}{\columnwidth}
        \centering
        \includegraphics[width=\columnwidth]{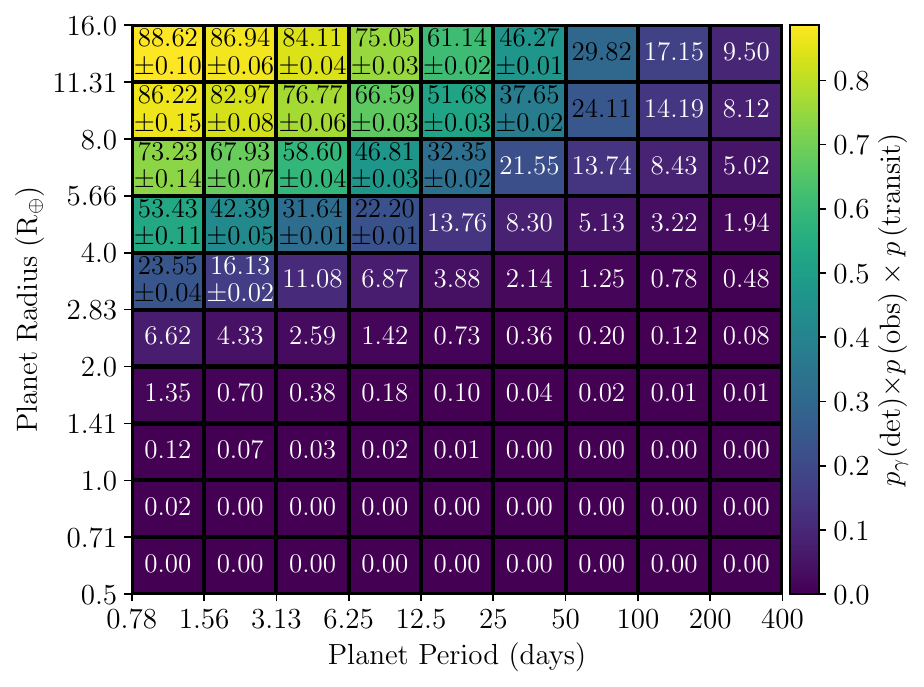}
        \caption{F type dwarf stars}
        \label{fig:sensitivity_F}
    \end{subfigure}
    \begin{subfigure}{\columnwidth}
        \centering
        \includegraphics[width=\columnwidth]{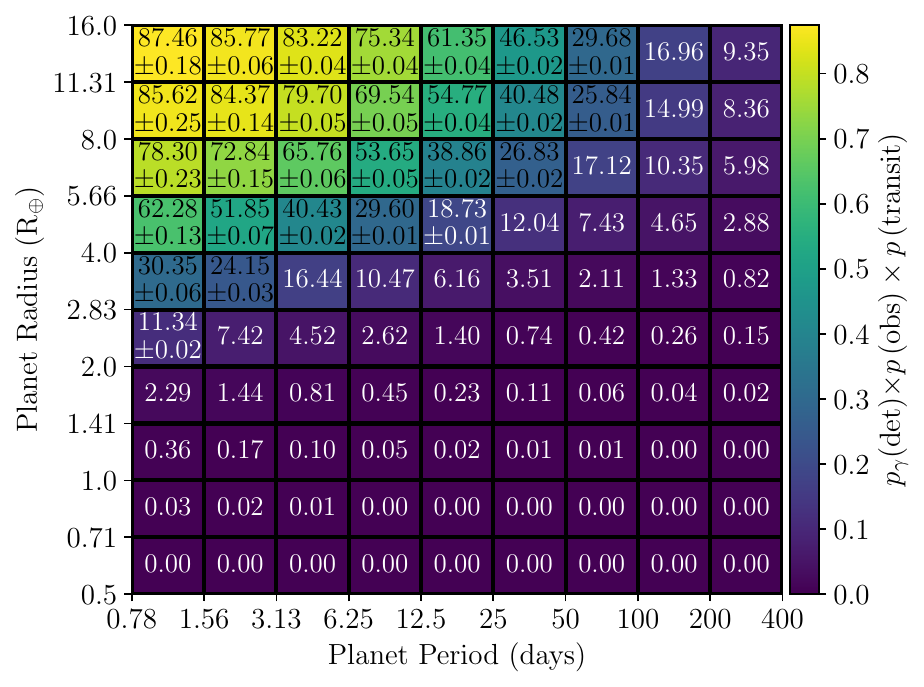}
        \caption{G type dwarf stars}
        \label{fig:sensitivity_G}
    \end{subfigure}
    \begin{subfigure}{\columnwidth}
        \centering
        \includegraphics[width=\columnwidth]{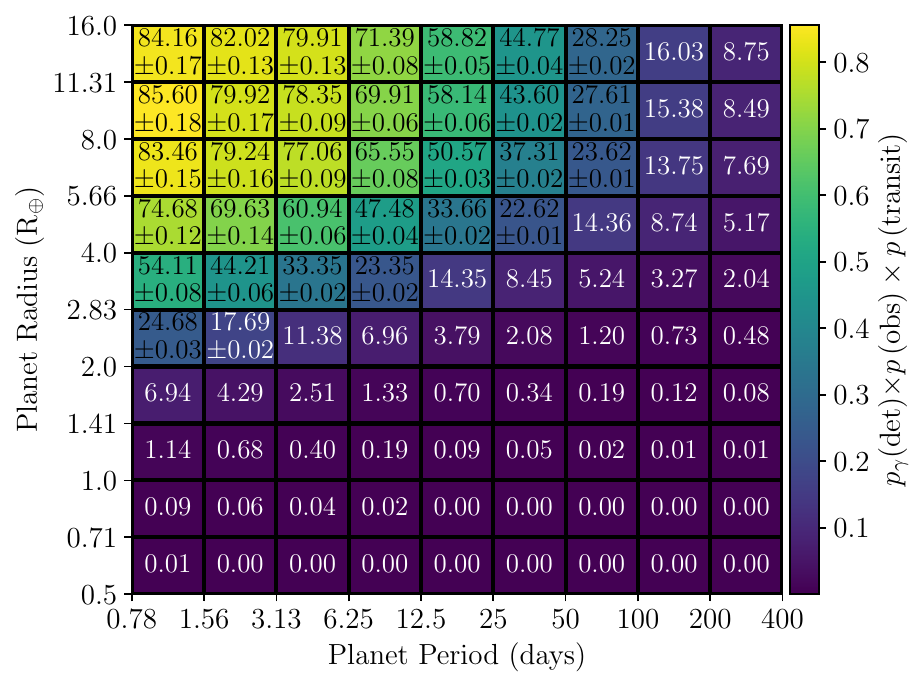}
        \caption{K type dwarf stars}
        \label{fig:sensitivity_K}
    \end{subfigure}
    \begin{subfigure}{\columnwidth}
        \centering
        \includegraphics[width=\columnwidth]{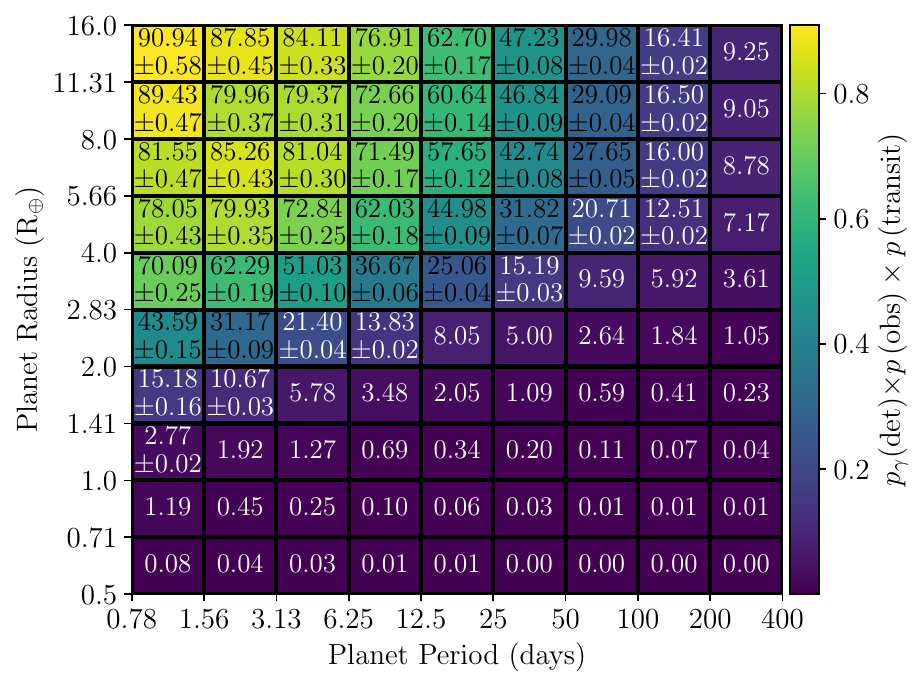}
        \caption{M type dwarf stars}
        \label{fig:sensitivity_M}
    \end{subfigure}
    \caption[Sensitivity maps for spectral types]{Sensitivity maps arranged by spectral type showing the probability of a transiting exoplanet being observed by \tess\ and then detected in the SPOC FFI lightcurves for Years 1 and 3. Numbers in each grid cell are the percentage sensitivity values. Where no uncertainties are given the uncertainty is 0\% to two decimal places.}
    \label{fig:sensitivity_types}
\end{figure*}

\section{Yield by spectral type}
\label{section:appendix_yield}

\begin{figure*}
    \centering
    \begin{subfigure}{\columnwidth}
        \centering
        \includegraphics[width=\columnwidth]{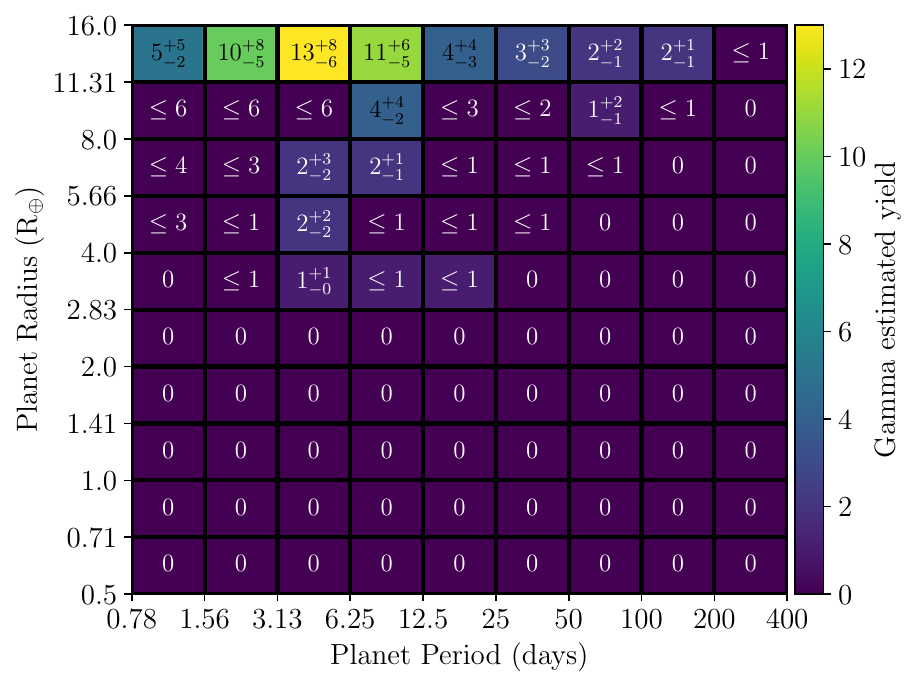}
        \caption{A type dwarf stars}
        \label{fig:yields_A}
    \end{subfigure}
    \begin{subfigure}{\columnwidth}
        \centering
        \includegraphics[width=\columnwidth]{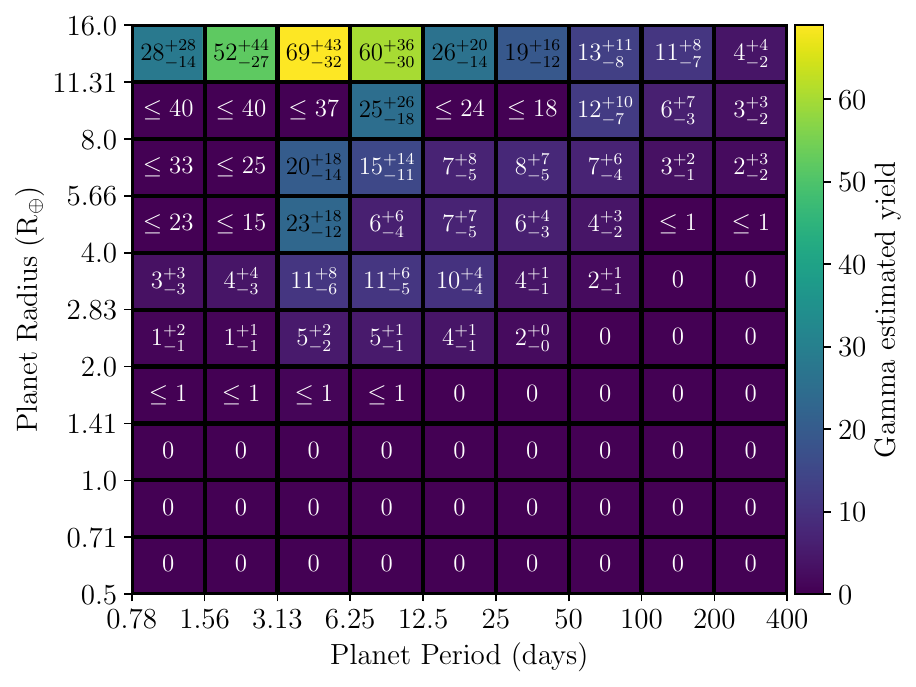}
        \caption{F type dwarf stars}
        \label{fig:yields_F}
    \end{subfigure}
    \begin{subfigure}{\columnwidth}
        \centering
        \includegraphics[width=\columnwidth]{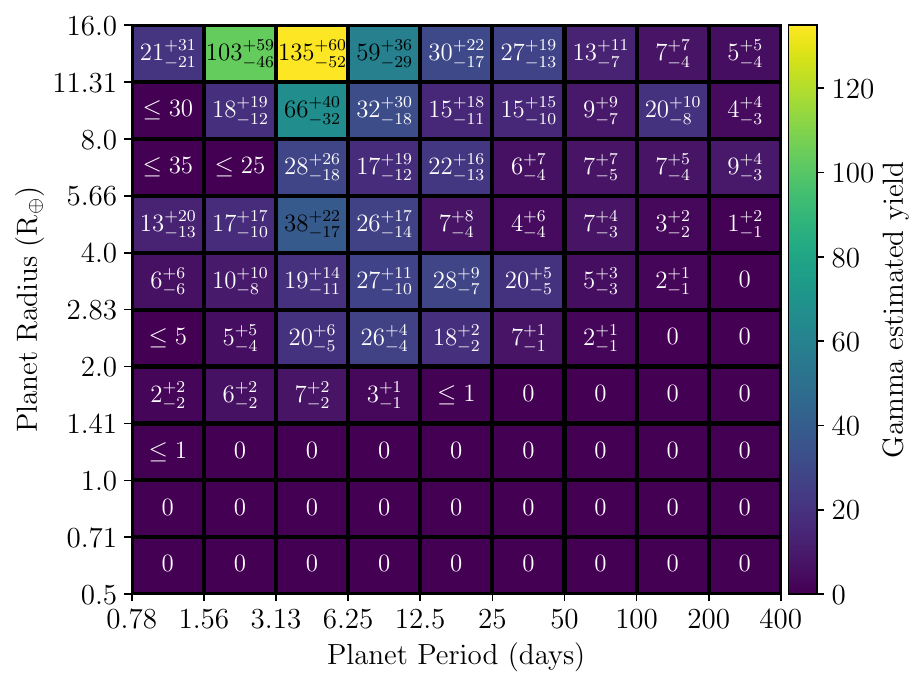}
        \caption{G type dwarf stars}
        \label{fig:yields_G}
    \end{subfigure}
    \begin{subfigure}{\columnwidth}
        \centering
        \includegraphics[width=\columnwidth]{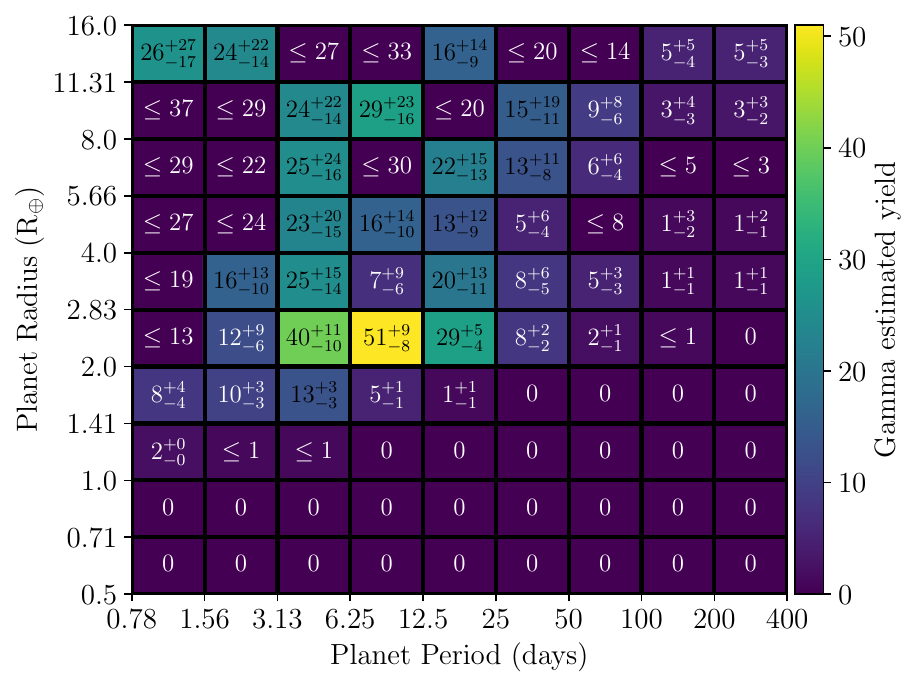}
        \caption{K type dwarf stars}
        \label{fig:yields_K}
    \end{subfigure}
    \begin{subfigure}{\columnwidth}
        \centering
        \includegraphics[width=\columnwidth]{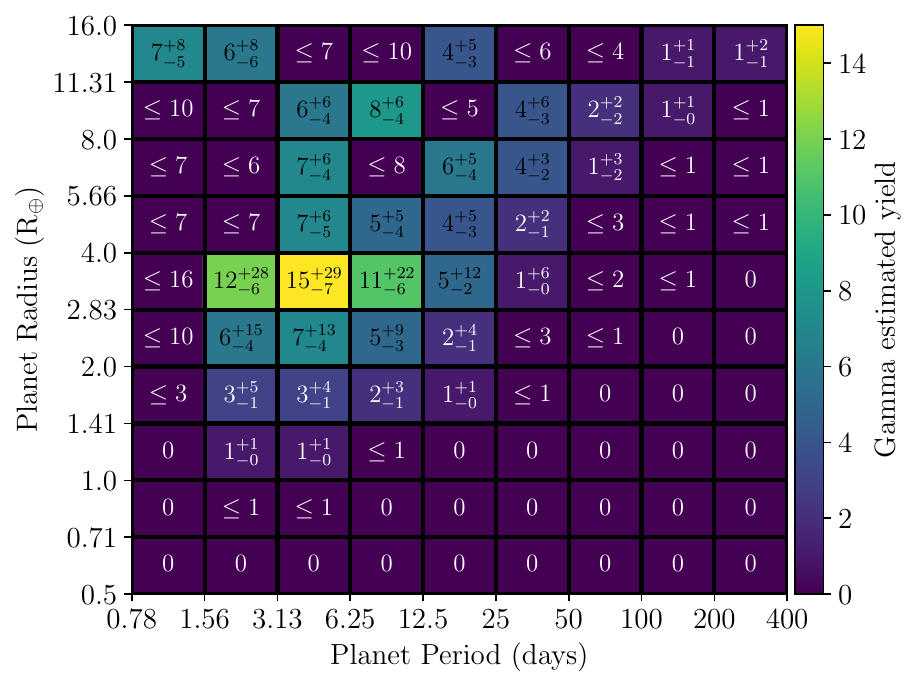}
        \caption{M type dwarf stars}
        \label{fig:yields_M}
    \end{subfigure}
    \caption[Yield maps for spectral types]{Predicted yields of exoplanets from the Years 1 and 3 \tess\ SPOC FFI lightcurves divided by spectral type.}
    \label{fig:yield_types}
\end{figure*}

\bsp	
\label{lastpage}
\end{document}